\documentclass[aps,prl,twocolumn,showpacs,superscriptaddress]{revtex4-2}

\usepackage{amsmath}
\usepackage{graphicx}
\usepackage{lmodern}
\usepackage{amsmath}

\usepackage{color}
\usepackage{amssymb}
\usepackage{bm}
\usepackage{braket}
\usepackage{mathtools}
\usepackage{slashed}
\usepackage{bibunits}

\newcommand{\mr}{moir\'{e} }

\newcommand{\tr}{\text{Tr}}

\DeclareMathOperator{\Tr}{Tr}

\usepackage[dvipsnames]{xcolor}
\usepackage[colorlinks=true]{hyperref}
\hypersetup{
    colorlinks=true,
    linkcolor=blue,
    citecolor=blue,
    urlcolor=blue
}

\makeatletter
\def\maketitle{
\@author@finish
\title@column\titleblock@produce
\suppressfloats[t]}
\makeatother
\begin{document}

\title {Exact pair density wave in  topological moire flat bands and universal superfluid stiffness 
}
\date{\today}
\author{Zhengzhi Wu}
\thanks{These two authors contributed equally to this work.}
\affiliation{Rudolf Peierls Centre for Theoretical Physics, Parks Road, Oxford, OX1 3PU, UK}
\author{Ming-Rui Li}
\thanks{These two authors contributed equally to this work.}
\affiliation{Institute for Advanced Study, Tsinghua University, Beijing 100084, China}
\author{Hong Yao}
\email{yaohong@tsinghua.edu.cn}
\affiliation{Institute for Advanced Study, Tsinghua University, Beijing 100084, China}

\begin{abstract}
Pair-density waves (PDWs) are unconventional superconducting states beyond the BCS paradigm, but identifying them unambiguously in microscopic lattice models is generally challenging due to various competing orders.
In particular, an exact realization of PDW in interacting models with topological bands remains elusive. Here we construct an interacting model on a twisted bilayer checkerboard lattice (TBCB) featuring topological flat bands, 
and rigorously show that an exact PDW ground state can be induced by quantum geometric nesting (QGN), owing to its momentum-space nonsymmorphic symmetry. We further prove that the superfluid stiffness of {\it any} nondegenerate QGN superconductor (including the PDW studied here and uniform superconductors) obeys the universal relation $D_s=2N_{\rm flat}\nu(1-\nu)m_{\rm pair}^{-1}$, where $N_{\rm flat}$, $\nu$, and $m_{\rm pair}$ are the number of flat bands, flat band filling, and the two-particle effective mass, respectively.
Our results 
reveal a rich interplay among PDW order, quantum geometry, and band topology. 
\end{abstract}
\maketitle

{\bf Introduction:} Pair-density waves (PDWs) are unconventional superconductors whose order parameters periodically oscillate in real space with a vanishing spatial average \cite{annurev}.  They are not only intriguing novel phases of matter distinct from uniform superconductors but also parent states for even more exotic phases and quantum critical points, such as  charge-$4e$ or charge-$6e$ superconductivity \cite{Berg2009,Zhou2022} 
and supersymmetric quantum criticality \cite{PhysRevLett.114.237001,PhysRevLett.118.166802}. Despite accumulating experimental evidence \cite{Hamidian2016,Ruan2018,Edkins2019,Du2020CupratePDW,PhysRevX.11.011007,Liu2021TMDPDW,PhysRevB.106.174510,lee2023,Chen2021,Deng2024KagomePDW,Gu2023,Zhao2023,Liu2023,Kong2025IronPDM,Wang2026MoireCPDM}  and sustained theoretical interest in PDWs and related finite-momentum pairing states \cite{PhysRev.135.A550,LO,PhysRevLett.99.127003,PhysRevB.79.064515,Berg_2009,Yang_2009,PhysRevB.82.041102,PhysRevLett.105.146403,PhysRevB.85.035104,PhysRevX.4.031017,PhysRevLett.114.197001,PhysRevB.98.224501,PhysRevX.9.021047,VenderleyKim2019,PhysRevLett.122.167001,PhysRevLett.125.167001,Huang2022HolsteinHubbardPDW,PhysRevLett.129.167001,PhysRevLett.130.026001,PhysRevLett.131.016002,Setty2023,PhysRevB.107.045122,PhysRevLett.130.126001,PhysRevLett.131.016001,PhysRevLett.131.026601,PhysRevB.110.094515,PhysRevLett.133.176501,PhysRevB.112.L140505,Wang2025,Sun2025,barlas2025quantumgeometryinducedkekule,Zhu2025OrbitalFFLOTwistedWSe2,wy3f-hgr9,Chen2026}, many fundamental properties of PDWs remain elusive. One such poorly understood physical quantity is the PDW superfluid stiffness, which can be anomalously small, or even nonpositive in mean-field treatments, despite a nonzero pairing gap \cite{PhysRevB.98.224501,Wang2026,ylr3-d2h1}. In uniform superconductors, the geometric contribution to the superfluid stiffness has been extensively studied in flat-band and moir\'{e} settings \cite{Peotta2015,PhysRevB.94.245149,PhysRevB.98.220511,PhysRevLett.123.237002,PhysRevB.101.060505,PhysRevB.106.014518,herzogarbeitman2022,PhysRevLett.128.087002}. More specifically, lower and topology-based bounds have been established for uniform superconductors in topologically nontrivial flat bands \cite{Peotta2015,PhysRevLett.124.167002,PhysRevLett.128.087002,PhysRevB.107.L201106,5hxp-pjhl,gw85-5r92}. In contrast, analogous bounds for PDW stiffness remain unknown.

\begin{figure}[t]
    \includegraphics[width=1\linewidth]{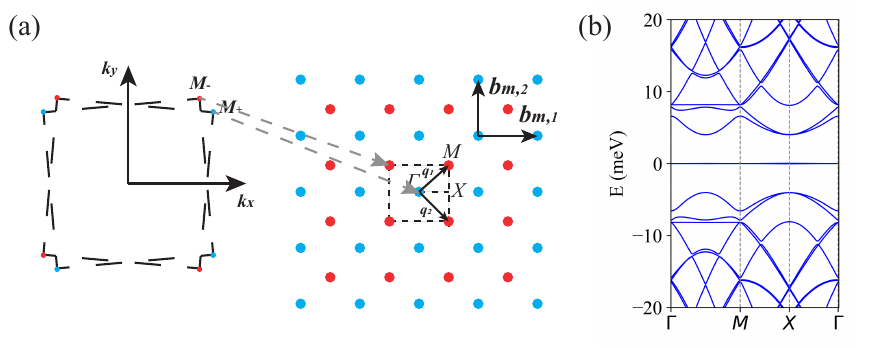}
\caption{(a) Moir\'{e} Brillouin zone of the TBCB model, illustrating the M-point moir\'{e} structure and the finite momentum transfer relevant for the nonsymmorphic symmetry. (b) Band structure in the chiral limit. Two exact flat bands (per spin) lie at zero energy, while the dispersive bands are symmetric with respect to the flat bands owing to the chiral symmetry. 
}
    \label{fig:minibz}
\end{figure}

To rigorously investigate the properties of PDWs, we construct exact PDW ground states using the recently proposed framework of quantum geometric nesting (QGN) \cite{PhysRevX.14.041004}. QGN provides a route to identify the leading instabilities and construct exact symmetry-breaking ground states in flat-band systems.
While an engineered model hosting QGN-compatible PDWs in topologically trivial bands was  proposed in \cite{PhysRevX.14.041004}, it remains unknown whether QGN of PDW orders can arise naturally in a physical setup---particularly within topologically nontrivial flat bands--- and, if so, whether it enables the construction of exact PDW ground states.

Here we propose that a promising physical platform for realizing such novel physics is the recently proposed class of $\mathbf{M}$-point or $\mathbf{M}$-valley twisted \mr systems \cite{Calugaru2025,Lei2025MValley}. In contrast to twisted bilayer graphene and other moir\'{e} platforms rooted in valleys at $\mathbf{\Gamma}$ or $\pm\mathbf{K}$ points \cite{BistritzerMacDonald2011,Cao2018Superconductivity,Cao2018Correlated,PhysRevLett.122.086402,PhysRevLett.123.237002,PhysRevB.101.060505,PhysRevB.103.205415,Andrei2021Marvels,Kennes2021Simulator,PhysRevLett.129.047601,PhysRevLett.132.096602}, these systems feature monolayer low-energy states near the time-reversal symmetric  $\mathbf{M}$ point (modulo a nonzero reciprocal vector) \cite{Calugaru2025,Lei2025MValley}. A fascinating property of $\mathbf{M}$-point moir\'{e} systems is the emergence of the momentum-space nonsymmorphic symmetry, analogous to fractional-flux Hofstadter physics and its projective translation algebra \cite{Brown1964,Zak1964,Hofstadter1976,PhysRevLett.49.405,PhysRevLett.111.185301,Chen2022,PhysRevLett.130.256601,PhysRevLett.130.236601}, but without requiring an external magnetic field. While previous studies have shown that this non-symmorphic symmetry could suggest the possibility of insulators or simulating a Luttinger liquid in two-dimensional materials \cite{Calugaru2025,li2025emergentinteractingphasesstrong,Ingham2025MValleyBilayers,Bao2025MValleyMXenes,DeBeule2025MValleyInteractions,mixeddimensionalquantummontecarlo,vasiliou2026}, several fundamental questions persist. It is unknown whether this momentum-space non-symmorphic symmetry is a universal feature of $\mathbf{M}$-point moiré systems, or simply whether there are other $\mathbf{M}$-point moiré systems with the non-symmorphic symmetry besides the two examples proposed in \cite{Calugaru2025}, particularly in settings hosting topologically nontrivial flat bands. Moreover, the broader physical consequences of such nonsymmorphic symmetries—beyond their connection to Luttinger-liquid behavior and insulators—remain largely unexplored.

Here we address these issues by introducing a simple twisted bilayer checkerboard lattice (TBCB) model. This model builds on the checkerboard-lattice quadratic band crossing \cite{PhysRevLett.103.046811} and its twisted-bilayer realization with topologically nontrivial flat bands \cite{PhysRevResearch.4.043151,PhysRevB.110.165142,LiWuYaoUnpublished}. Viewed as an $\mathbf{M}$-point \mr system, it realizes momentum-space nonsymmorphic symmetry 
\cite{Calugaru2025}. We demonstrate that the TBCB model provides a natural physical setting for the QGN of PDW states. Specifically, the combination of nonsymmorphic symmetry and time-reversal symmetry strictly enforces the required QGN structure for PDW order. Building on this symmetry-enforced QGN, we explicitly construct realistic local interactions for which the PDW state is an exact ground state---constituting, to our knowledge, the first exactly solvable PDW ground state in topological \mr flat bands. We further prove that the superfluid stiffness of {\it any}  nondegenerate QGN superconductors  universally takes the form:  $D_s=2N_{\rm flat}\nu(1-\nu)m_{\rm pair}^{-1}$ in the thermodynamic limit, which was a conjecture in previous studies of QGN superconductors \cite{gw85-5r92}. 
Finally, using a mean-field analysis, we elucidate the interplay among superfluid stiffness, quantum geometry, and band topology, and demonstrate that the superfluid stiffness in our model admits a topological lower bound.

{\bf Model and Symmetries:} 
As a model with the minimal number of topologically stable valleys at the $M$-point, the monolayer checkerboard lattice hosts a single quadratic band touching (QBT) described by the single-particle Hamiltonian \cite{PhysRevLett.103.046811}: 
\begin{align}
    h(\mathbf{k})=t_z(k_x^2-k_y^2)\sigma_z+t_xk_xk_y\sigma_x,
\end{align}
where $\sigma_i$ acts on the A/B sublattice,  and $\mathbf{k}$ is measured relative to the valley $\frac{1}{a}(\pi,\pi)$ with $a$ the monolayer lattice constant. To realize topological flat bands, we construct the Bistritzer-MacDonald (BM) continuum model of the TBCB system \cite{PhysRevResearch.4.043151}, where each layer is rotated by $\frac{l\theta}{2}$ relative to the AA stacking configuration, with $l=\pm$ labeling the top and bottom layers. The kinetic energy takes the form: 
\begin{align}
\hat{H}_0=\sum_{\mathbf{Q}_l,\mathbf{Q}_l^{\prime}}\sum_{\mathbf{k}}\hat{f}_{\mathbf{Q}_l}^{\dagger}(\mathbf{k})h_{\mathbf{Q}_l,\mathbf{Q}_l^{\prime}}(\mathbf{k})\hat{f}_{\mathbf{Q}_l^{\prime}}(\mathbf{k}),
\end{align}
where the fermion operator $\hat{f}_{\mathbf{Q}_l}(\mathbf{k})=\hat{f}_{\mathbf{Q}_l,\sigma,l,s}(\mathbf{k})$ carries the momentum $\mathbf{k}$, and $s=\pm1$ represents the spin-$\frac{1}{2}$ degrees of freedom. The single-particle Hamiltonian $h_{\mathbf{Q}_l,\mathbf{Q}_l^{\prime}}(\mathbf{k})$ takes the form of a tight-binding model in momentum space. The primitive \mr reciprocal momenta are $\mathbf{b}_{m,1}=a_m^{-1}(2\pi,0)$ and $\mathbf{b}_{m,2}=a_m^{-1}(0,2\pi)$, where the \mr lattice constant is $a_m=\frac{a}{2\sin{\theta/2}}$.  $\mathbf{Q}_{l}=\mathbb{Z}\mathbf{b}_{m,1}+\mathbb{Z}\mathbf{b}_{m,2}-\mathbf{M}_{l}$ denotes the reciprocal lattice generated by the \mr reciprocal vectors and shifted by the valley momentum $\mathbf{M}_l$ of the layer $l$ ($\mathbf{M}_+=0,\mathbf{M}_-=-\frac{1}{a_m}(\pi,\pi)$), as illustrated in Fig.~\ref{fig:minibz}(a). The \mr potential couples fermions whose momenta differ by a reciprocal momentum, reflecting the breaking of the monolayer translation symmetry by the \mr potential. Written explicitly, the single-particle \mr Hamiltonian takes the form \footnote{Here we adopt the same small twist-angle limit $\theta\ll1$ as in \cite{Calugaru2025}, in which $\theta$ enters into $h^0_{\mathbf{Q}_l,l}(\mathbf{k})$ only through $\mathbf{Q}_l$. }:
\begin{align}
        &h_{\mathbf{Q}_l,\mathbf{Q}_{l'}^{\prime}}(\mathbf{k})=\delta_{\mathbf{Q}_l,\mathbf{Q}_l^{\prime}}h^0_{\mathbf{Q}_l,l}(\mathbf{k})+h_{\mathbf{Q}_l,\mathbf{Q}_{l'}^{\prime}}^{t}(\mathbf{k}),\nonumber\\
         &h^0_{\mathbf{Q}_l,l}(\mathbf{k})=h(\mathbf{k}-\mathbf{Q}_l),\quad h_{\mathbf{Q}_l,\mathbf{Q}_{l'}^{\prime}}^{t}(\mathbf{k})=\delta_{l,-l^{\prime}}\sum_{i=1,2}\delta_{\mathbf{Q}_{l'}^{\prime},\mathbf{Q}_l\pm\mathbf{q}_i}T_i,\nonumber\\
        &T_1 = \begin{pmatrix} w_{AA} & w_{AB} \\ w_{AB} & w_{AA} \end{pmatrix}, \quad T_2 = \begin{pmatrix} w_{AA} & -w_{AB} \\ -w_{AB} & w_{AA} \end{pmatrix},
\end{align}
where $\mathbf{q}_1=\frac{1}{a_m}(\pi,\pi)$ and $\mathbf{q}_2=\frac{1}{a_m}(\pi,-\pi)$. $h^0_{\mathbf{Q}_l,l}(\mathbf{k})$ and $h_{\mathbf{Q}_l,\mathbf{Q}_{l'}^{\prime}}^{t}(\mathbf{k})$ denote the intra-layer kinetic energy and inter-layer tunneling, respectively, while the matrices $T_i$ act in the sublattice subspace. The parameters $w_{AA}$ and $w_{AB}$ represent the intra- and inter-sublattice hopping amplitudes. In the following discussion, we focus on the chiral limit ($w_{AA}=0$) at magic angles, where two exact topological flat bands (per spin) emerge at zero energy (Fig.~\ref{fig:minibz}). There is a band gap $\Delta_g$ between the flat bands and the dispersive bands, as illustrated in Fig.~\ref{fig:minibz}. The nontrivial band topology is characterized by the nonzero winding number of the Wilson loop, yielding $e_2=2$ for the two flat bands \cite{PhysRevResearch.4.043151}.

The emergence of QGN is fundamentally rooted in the symmetries of the TBCB model. First, the Hamiltonian $\hat{H}_0$ exhibits a momentum-space nonsymmorphic symmetry that effectively exchanges the two layers \cite{SupMat}:
\begin{equation}
    \tilde{M}_z\hat{f}_{\mathbf{Q}_l,l}(\mathbf{k})\tilde{M}_z^{-1}=\hat{f}_{\mathbf{Q}_l-\mathbf{q}_1,-l}(\mathbf{k}-\mathbf{q}_1).
    \label{nonsym}
\end{equation}
$\tilde{M}_z$ is the combination of two symmetries with more transparent physical meanings: $\tilde{M}_z=M_{x/y}\circ \tilde{C}_{2y/x}$, where $M_{x/y}$ and $\tilde{C}_{2y/x}$ are the mirror symmetries and layer-exchanging $\pi$-rotations:
\begin{equation}
\begin{aligned}
     &M_{x/y} \hat{f}_{\mathbf{Q}_l,l}(\mathbf{k}) M_{x/y}^{-1}= \sigma_{z} \hat{f}_{M_{x/y} \mathbf{Q}_l,l}(M_{x/y}\mathbf{k}), \\
&\tilde{C}_{2x/y} \hat{f}_{\mathbf{Q}_l,l}(\mathbf{k}) \tilde{C}_{2x/y}^{-1} = \sigma_z  \hat{f}_{ C_{2x/y} \mathbf{Q}_l - \mathbf{q}_{1},-l}(C_{2x/y} \mathbf{k} - \mathbf{q}_{1}),
\end{aligned}
\end{equation}
where $\tilde{C}_{2x/y}$ also acts nonsymmorphically in momentum space.  Nevertheless, we refer to the symmetry $\tilde{M}_z$ as an ``intrinsic momentum-space nonsymmorphic symmetry'', as it shifts the momentum $\mathbf{k}$ by a fraction of the reciprocal momentum without involving any additional point-group operations. The distinction by ``intrinsic" is not only motivated by the fact that only intrinsic nonsymmorphic symmetries like $\tilde{M}_z$ can induce QGN of PDW orders, as  we elaborate below, but also because certain nonsymmorphic symmetries that involve additional point-group transformations can be reduced to symmorphic ones through gauge transformations. For instance, this occurs in a topologically trivial twisted square lattice model with a space group $P$-$4m2$ \cite{bao2026}. In contrast, $\tilde{M}_z$ cannot be made symmorphic by any gauge transformations.

Physically, $\tilde{M}_z$ enforces a finite-momentum ($\mathbf{q}_1$) ``nesting" of the Bloch wavefunctions: $  u_{\mathbf{Q}_l,l,\sigma,s}^n(\mathbf{k})\propto \delta_{l+l^{\prime},0}\delta_{\mathbf{Q}_{l^{\prime}}+q_1,\mathbf{Q}_l}u^n_{\mathbf{Q}_{l^{\prime}},l^{\prime},\sigma,s}(\mathbf{k}-\mathbf{q}_1)$. As a result, two normalized Bloch states within the band labeled by $n$, separated by momentum $\mathbf{q}_1$, are necessarily connected by a unitary matrix. This constitutes a form of geometric nesting, because it arises from the structure of the Bloch wave functions rather than from conventional Fermi surface nesting. Furthermore, in the chiral limit, $\hat{H}_0$ preserves an anti-unitary chiral symmetry defined by $\hat{f}_{\mathbf{Q}_l}(\mathbf{k})\rightarrow \sigma_y \hat{f}_{\mathbf{Q}_l}^{\dagger}(\mathbf{k})$. Consequently, this chiral symmetry renders the dispersive energy spectrum perfectly symmetric with respect to the zero-energy flat bands.

{\bf Exactly Solvable Pair-Density Wave and QGN:} In this section, we construct interactions $\hat{V}$ such that the full Hamiltonian $\hat{H}=\hat{H}_0+\hat{V}$ hosts an exact PDW ground state utilizing the QGN framework~\cite{PhysRevX.14.041004}. We focus on the filling regime in which the flat bands are partially occupied in the absence of interactions, and on the limit where the interaction scale $U$ is much smaller than the band gap $\Delta_g$. This nevertheless corresponds to the strong-coupling limit, since the (active) band width is exactly zero and the resulting PDW ground state is driven entirely by interactions, which is in spirit similar to the FQH systems.  In this regime, the dispersive bands are effectively ``frozen", and the exact ground state of the flat-band-projected Hamiltonian $\hat{H}_{\text{eff}}=\hat{P} \hat{H} \hat{P}$ becomes an asymptotically exact ground state of the full Hamiltonian $\hat{H}$ as $U/\Delta_g\rightarrow0$. The operator $\hat{P}$ projects onto the subspace spanned by $\{|\psi_i\rangle=|i\rangle\otimes|0\rangle\}$, where $\{|i\rangle\}$ denotes the basis of the flat-band Fock space, and $|0\rangle$ is the unique state wherein the lower (upper) dispersive bands are completely filled (empty). Consequently,  we focus on the effective Hamiltonian $\hat{H}_{\text{eff}}=\hat{P} \hat{V} \hat{P}$ in the following discussions.

Concretely, we construct the following local interaction which hosts an exact PDW ground state:
\begin{equation}
    \hat{V}=U\int d^2\mathbf{x} \left((\hat{S}^{(0)}(\mathbf{x}))^2+\frac{1}{3}\left[\sum_i(\hat{S}^{(i)}(\mathbf{x}))^2\right]\right),
    \label{eq:interaction_hamiltonian_pdw}
\end{equation}
where $U>0$, $\hat{S}^{(0)}(\mathbf{x})=\hat{f}^{\dagger}(\mathbf{x})\tau_z\hat{f}(\mathbf{x})$, and $\hat{S}^{(i)}(\mathbf{x})=\hat{f}^{\dagger}(\mathbf{x})\tau_z\sigma_ys^{i}\hat{f}(\mathbf{x})$. The real space fermion operator is defined as: $\hat{f}^{\dagger}_{l,\sigma,s}(\mathbf{x})=\frac{1}{\sqrt{S}}\sum_{\mathbf{k}}\sum_{\mathbf{G}}\hat{f}_{\mathbf{G}-\mathbf{M}_l,l,\sigma,s}^{\dagger}(\mathbf{k})e^{-i(\mathbf{k}-\mathbf{G})\cdot \mathbf{x}}$
where $\mathbf{G}=\mathbb{Z}\mathbf{b}_{m,1}+\mathbb{Z}\mathbf{b}_{m,2}$ represents the reciprocal momentum and $S$ denotes the total area of the system. 
An exact ground state of $\hat{H}_{\text{eff}}$, exhibiting off-diagonal long-range order (ODLRO), is given by the PDW state in the generalized $\eta$-pairing form \cite{PhysRevLett.63.2144}:
\begin{equation}
\begin{aligned}
    |\eta_N\rangle&=\left((\tilde{\eta}^\dagger)^p|\text{vac}\rangle\right)\otimes |0\rangle,\qquad p=N/2,\\
    \tilde{\eta}^{\dagger}&=\hat{P}\left(\int d^2\mathbf{x} \hat{f}^{\dagger}(\mathbf{x})(\tau_xs_y)(\hat{f}^{\dagger}(\mathbf{x}))^Te^{i\mathbf{q}_1\cdot\mathbf{x}}\right)\hat{P}\\
    &=\sum_{\mathbf{k},m\in\text{flat}}e^{-i\Theta_{\mathbf{k},m}}\hat{\gamma}_{m,s}^{\dagger}(\mathbf{k})(s_y)_{s,s^{\prime}}\hat{\gamma}^{\dagger}_{m,s^{\prime}}(-\mathbf{k}+\mathbf{q}_1)
\end{aligned}
\end{equation}
  where $N$ is the total number of electrons in the flat bands, and $\gamma_{m,s}^{\dagger}(\mathbf{k})$ creates a fermion in the band labeled by $m$.  Physically, $\tilde{\eta}^{\dagger}$  describes a singlet PDW order carrying a finite center-of-mass momentum $\mathbf{q}_1=\frac{1}{a_m}(\pi,\pi)$, and the phase factor $e^{-i\Theta_{\mathbf{k},m}}$ is fixed by the action of the symmetry $\tilde{M}_z\cdot T$, with $T=s_y K$ representing the time-reversal symmetry (TRS), on the Bloch wavefunctions:
  \begin{align}
u^{(m)}_{\mathbf{Q}_l,l,\sigma,s}(\mathbf{k}) = e^{i\Theta_{\mathbf{k},m}} (u^{(m)}_{-\mathbf{Q}_l+\mathbf{q}_1, -l,\sigma,-s}(-\mathbf{k}+\mathbf{q}_1))^*,    
\label{symmetry}
\end{align}
where $e^{i\Theta_{\mathbf{k},m}}$ depends exclusively on the band index $m$ and the moir\'{e} momentum $\mathbf{k}$ \footnote{Within degenerate subspaces, the symmetry relations generally involve unitary sewing matrices. The scalar phases used here and in Eqs.~(S8)--(S9) refer to a symmetry-adapted orbital basis, defined patchwise. The QGN condition is independent of this basis choice.}.

We can rigorously prove that the PDW state $|\eta_N\rangle$ is 
an exact ground state of the positive semi-definite Hamiltonian $\hat{H}_{\text{eff}}$, as detailed in the Supplemental Material \cite{SupMat}. The proof strategy is sketched as follows: $\hat{H}_{\text{eff}}$ is a positive semi-definite Hamiltonian: $\hat{H}_{\text{eff}}=U\int d^2\mathbf{x} \left((\tilde{S}^{(0)}(\mathbf{x}))^2+\frac{1}{3}\left[\sum_i(\tilde{S}^{(i)}(\mathbf{x}))^2\right]\right)$. Here, $\tilde{S}^{\mu}(\mathbf{x})$ denotes the projected $\hat{S}^{\mu}(\mathbf{x})$ involving only fermion operators within the flat-band subspace, which is related to the unprojected operator via $\hat{P}\hat{S}^{\mu}(\mathbf{x})\hat{P}=\tilde{S}^{\mu}(\mathbf{x})+\langle0|\hat{S}^{\mu}(\mathbf{x})|0\rangle$. Crucially, while projecting a four-fermion interaction generally generates additional quadratic Hartree-Fock (HF) type one-body terms that endow the flat bands a finite bandwidth of order $U$,  no such terms arise in $\hat{H}_{\text{eff}}$ due to the chiral and nonsymmorphic symmetries ($\tilde{M}_z$). The absence of one-body terms here goes beyond the ``uniform pairing condition" proposed in  QGN models with finite orbitals \cite{PhysRevB.94.245149,herzogarbeitman2022, PhysRevX.14.041004}. Since $\eta^{\dagger}$ satisfies QGN, as elaborated below, and obeys $[\eta^{\dagger},\hat{S}^{\mu}(\mathbf{x})]=0$,  the commutator of the projected operators also vanishes: $[\tilde{\eta}^{\dagger},\tilde{S}^{\mu}(\mathbf{x})]=0$, with the proof deferred to the End Matter. Consequently, the PDW state $|\eta_N\rangle$ is a zero-energy state, and hence a ground state, of the positive semi-definite $\hat{H}_{\text{eff}}$.  

 An order parameter is defined to satisfy the QGN condition if its form factor is strictly block-diagonal in the energy-band basis, ensuring no matrix elements couple the flat-band subspace to the dispersive bands. Crucially, the $\eta^{\dagger}$ in our model satisfies QGN due to the combined symmetry $\tilde{M}_z\cdot T$. The QGN structure of $\eta^{\dagger}$ becomes fully transparent when written in the band-basis:
\begin{equation}
    \eta^{\dagger} =\sum_{\mathbf{k},m,n}\gamma_{m,s}^{\dagger}(\mathbf{k})F_{m,s;n,s^{\prime}}(\mathbf{k})\gamma_{n,s^{\prime}}^{\dagger}(-\mathbf{k}+\mathbf{q}_1).
\end{equation}
 The pairing form factor $F_{m,s;n,s^{\prime}}(\mathbf{k})$ is determined by the overlap of the Bloch wavefunctions $u^{(n)}$:
\begin{align}
    &F_{m,s;n,s^{\prime}}(\mathbf{k})\nonumber\\
    &=\sum_{\mathbf{Q}_l,l,\sigma}(u_{\mathbf{Q}_l,l,\sigma,s}^{(m)}(\mathbf{k}))^*\left((s_y)_{s,s^{\prime}}\right)(u_{-\mathbf{Q}_l+\mathbf{q}_1,-l,\sigma,s^{\prime}}^{(n)}(-\mathbf{k}+\mathbf{q}_1))^*.
\end{align} 
The form factor is block-diagonalized, i.e. $F_{m,s;n,s^{\prime}}(\mathbf{k})=0$ for any $m \in \text{flat}, n \notin \text{flat}$ (and vice versa), since the two Bloch wavefunctions separated by the moir\'{e} momentum $\mathbf{q}_1$ satisfy Eq. \eqref{symmetry} as a consequence of the symmetry $\tilde{M}_z\cdot T$.

Finally, a critical remaining question is whether the PDW state $|\eta_N\rangle$ is a unique ground state. We exhaustively enumerated all candidate orders with QGN, which are the leading instabilities since their susceptibilities within the flat bands saturate the theoretical upper bound \cite{PhysRevX.14.041004}, and found that, among the associated symmetry-breaking states, the PDW state $|\eta_N\rangle$ is the unique ground state of $\hat{H}_{\text{eff}}$ at generic fillings $\nu$ \cite{SupMat}. The sole exception occurs at $\nu=1/2$ (half-filling of the flat bands), where a quantum anomalous Hall (QAH) state $|\Psi\rangle$ becomes exactly degenerate with $|\eta_N\rangle$ \cite{SupMat}. However, this accidental degeneracy can be lifted by introducing a minimal perturbation which selects $|\eta_N\rangle$ as the unique exact ground state of $\hat{H}_{\text{eff}}+\hat{P}\Delta \hat{V}\hat{P}$:
\begin{equation}
\begin{aligned}
  &\Delta \hat{V}=V\int d^2x \left[(S^{(4)}(x))^2+(S^{(5)}(x))^2\right], \forall V>0 \\
  &S^{(4)}(x)=f^{\dagger}(x)\tau_z\sigma_xf(x),S^{(5)}(x)=f^{\dagger}(x)\tau_z\sigma_zf(x).
\end{aligned}
\end{equation}
In summary, we expect the PDW state to be the unique ground state at generic fillings. At the sole exceptional filling, $\nu=1/2$, the perturbed Hamiltonian  uniquely selects the PDW ground state. These conclusions are further corroborated by exact-diagonalization (ED) calculations.


{\bf Exact Excitation Spectrum:}
The few-body excitation spectrum, including both the single-particle and two-particle sectors, is also exactly solvable in our model.  The excitation operator $\hat{\mathcal{C}}^c_{\mathbf{p}}$ with conserved charge $c$ and momentum $\mathbf{p}$ satisfies \cite{SupMat}:
\begin{align}
[\hat{{H}}_{\text{eff}},\hat{\mathcal{C}}^c_{\mathbf{p}}]\ket{\Psi}=E_c(\mathbf{p})\hat{\mathcal{C}}^c_{\mathbf{p}}\ket{\Psi},
\end{align}
where $\ket{\Psi}$ is the ground state. $\hat{\mathcal{C}}^c_{\mathbf{p}}$ and $E_c(\mathbf{p})$ are the eigen-operator and corresponding eigen-energy of a scattering matrix $\Gamma^{\mathbf{p}}$, which arises  from the commutator between $\hat{{H}}_{\text{eff}}$ and the few-body fermion operators  with momentum $\mathbf{p}$ (single-fermion for $c=\pm1$ and quadratic-fermion operators for $c=\pm2$ ). For instance, the scattering matrix of single-hole excitations is defined by: $ [\hat{H}_{\text{eff}}, \hat{\gamma}_{m,s}(\mathbf{p})]|\Psi\rangle = \sum_{n,s'} \Gamma_{m,s;n,s'}^{h,\mathbf{p}} \hat{\gamma}_{n,s'}(\mathbf{p})|\Psi\rangle$. This equation closes within the single-fermion space since the ground state $|\Psi\rangle$ is a zero mode of $\tilde{S}^{\mu}(\mathbf{x})$.

\begin{figure}[!t]
\centering
    \includegraphics[width=\linewidth]{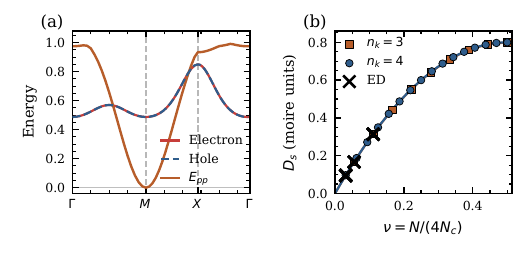}
\caption{Exact excitation spectra and PDW stiffness of the PDW ground state. (a) The single-particle excitation energy $E_{p/h}$ (the single electron and hole excitations are degenerate) and the spin-0 particle-particle excitation spectrum $E_{pp}$ evaluated along the high-symmetry line in the moir\'{e} Brillouin zone. The two-particle excitation is strictly gapless at the $M$ point, corresponding to the Goldstone mode. The apparent non-smoothness of the high-energy part of the plotted $E_{pp}$ branch reflects crossings with higher excitation branches. (b) Rescaled finite-size PDW stiffness with $N_C=n_k^2$ unit cells for $n_k=3,4$, which are represented as red and blue dots. The solid curve shows the analytical relation $D_s=2N_{\rm flat}\nu(1-\nu)m_{\rm pair}^{-1}$ with $m_{\rm pair}^{-1}$ extracted from the same-size two particle sector, and black crosses represent ED results of the stiffness with the same system size. For $n_k=3$, finite-mesh trace corrections cause small deviations from the solid curve. These corrections vanish in the thermodynamic limit.
}
    \label{fig:excitation_spectrum}
\end{figure}

We illustrated the single-particle and two-particle excitation spectra in Fig.~\ref{fig:excitation_spectrum}(a). The single-electron and hole excitations ($E_{p/h}$) are degenerate due to the chiral symmetry and are fully gapped. Remarkably, in contrast to other QGN lattice models with a finite number of orbitals \cite{herzogarbeitman2022,PhysRevX.14.041004}, our single-particle spectrum is dispersive rather than flat. Physically, the mobility of the single-particle excitations here originates from the nontrivial hopping structure in the $\mathbf{Q}_l$ lattice of the scattering matrix $\Gamma^{\mathbf{p}}$. This is an intrinsic feature of \mr systems, in contrast to QGN lattice models with a finite number of orbitals \cite{herzogarbeitman2022,PhysRevX.14.041004}, where the corresponding matrix is simply proportional to the identity.

Moreover, the spin-singlet two-particle excitation spectrum $E_{pp}(\mathbf{p})$ in Fig.~\ref{fig:excitation_spectrum}(a) is gapless at the $M$ point, corresponding to the Goldstone mode and providing a dynamical signature of the analytically predicted PDW ground state. Defining the dimensionless $\bar{\mathbf p}=(2\pi/|\mathbf b_{m,1}|)\mathbf p$ and $\bar{\mathbf Q}=(2\pi/|\mathbf b_{m,1}|)\mathbf Q$, the curvature of this charge-$2e$ mode defines the inverse pair mass at the PDW momentum,
\begin{equation}
    m_{\rm pair}^{-1}
    =
    \left.
    \frac{\partial^2 E_{pp}[(|\mathbf b_{m,1}|/2\pi)\bar{\mathbf p}]}
    {\partial \bar p_i^2}
    \right|_{\bar{\mathbf p}=\bar{\mathbf Q}}, i=x,y
\end{equation}
As we demonstrate in the next section, this two particle effective mass $ m_{\rm pair}^{-1}$ is directly connected to the superfluid stiffness, not only in our PDW model but in all QGN models with a superconducting ground state.

{\bf Exact superfluid stiffness and topological lower bound:}
First, we can prove that the zero temperature stiffness of any nondegenerate QGN superconductor \footnote{Here, “nondegenerate” means that the superconducting ground state is unique within a fixed-electron-number sector and the flat-band form factor $F(\mathbf{k})$ is full rank, as also assumed in Ref. \cite{gw85-5r92}, where Eq. \eqref{stiffness} was conjectured. If $F(\mathbf{k})$ is degenerate, some flat bands remain completely empty and unpaired, making the appropriate definitions of $N_{\text{flat},\nu}$ in Eq. \eqref{stiffness} model dependent. } universally takes the following form, with the details left to the Supplementary Material \cite{SupMat}:  
\begin{align}
    D_s
    =
    2N_{\rm flat}\nu(1-\nu)m_{\rm pair}^{-1},
    \label{stiffness}
\end{align}
where $D_s$ is the diagonal part of the stiffness tensor $[D_s]_{ij}=S^{-1}\partial_{\bar A_i}\partial_{\bar A_j}
E(\bar{\mathbf A})|_{\bar{\mathbf A}=0}$. $E(\bar{\mathbf A})$ is the ground state energy in the presence of a uniform small gauge field $\mathbf{A}$ and  $\bar{\mathbf A}=(2\pi/|\mathbf b_{m,1}|)\mathbf A$ is the dimensionless gauge field. $\nu,S$ are the flat-band electron filling and total number of unit-cells respectively. Here $m_{\rm pair}^{-1}$ is understood as the second order derivative at the center-of-mass pairing momentum  of the lowest spectrum in the two-particle sector in a general QGN superconductor. We want to remark that  the relation Eq.\eqref{stiffness} was previously shown to hold in several QGN superconductor models \cite{tormareview,PhysRevB.98.220511,PhysRevB.106.014518,herzogarbeitman2022,PhysRevX.14.041004,gw85-5r92} and was conjectured to hold generally for QGN superconductors \cite{gw85-5r92}. Here, we provide a  proof of this relation based on the $\eta$-pairing structure of the superconducting ground states of QGN models, thereby establishing the conjecture.

In our PDW model, we extract the effective mass $m_{\rm pair}^{-1}\simeq0.40$, as is illustrated  in Fig.~\ref{fig:excitation_spectrum}(b).  Since there are 
$N_{\rm flat}=4$ flat bands in total, this gives the exact PDW stiffness as
$D_s\simeq3.20\nu(1-\nu)$. 
To have a physical understanding of the quantum geometric and topological origin of this exact stiffness, we present a
complementary mean-field estimate for the PDW stiffness and derive a topological lower bound. We use a BdG approximation for the PDW state in the presence of a small gauge field $\mathbf{A}$, and the BdG ground state coincides with the exact PDW ground state at zero twist $\mathbf{A}=0$.
As detailed in the Supplemental Material, the mean-field stiffness is  determined by a quantum geometric contribution proportional to the Fubini-Study metric $g_{ij}(\mathbf{k})$ of the flat bands\cite{SupMat}:
\begin{equation}
\begin{aligned}
      D_s^{\rm MF}
      &=\frac{1}{2}\sum_i[D_s^{\rm MF}]_{ii}\\
      &=|\Delta_0|\sqrt{\nu(1-\nu)}
      \int \frac{d^{2}\mathbf{k}}{(2\pi)^{2}}
      \sum_i\operatorname{Tr}[g_{ii}(\mathbf{k})],
\end{aligned}
\end{equation}
 where $|\Delta_0|$ is the amplitude of the PDW order parameter. As $D_s^{\rm MF}$ is governed by the quantum metric, it obeys the topological lower bound $\int \frac{d^{2} \mathbf{k}}{(2 \pi)^{2}} \sum_i\operatorname{Tr}[g_{ii}(\mathbf{k})]\geq \frac{1}{\pi}e_2$, where $e_2=2$ is the winding number of the Wilson loop associated with the two flat bands (per spin). In order to obtain a lower bound, we associate the lowest single-particle excitation in the exact spectrum (Fig.~\ref{fig:excitation_spectrum}(a)) with the Bogoliubov quasiparticle energy $E_s = \sqrt{\mu^2 + \Delta_0^2}$, which in turn yields the following topological lower bound: $D_s^{\rm MF} \geq 2E_s \nu(1-\nu)e_2/\pi \approx 0.62\nu(1-\nu)$.  

This topological lower bound is reminiscent of that for uniform s-wave superconductivity in twisted bilayer graphene \cite{PhysRevLett.124.167002}. We emphasize that quantum geometric nesting is a necessary condition for both bounds, although it is enforced by different symmetries in the two settings: momentum space nonsymmorphic symmetry and TRS for the PDW here, and TRS and  $U(1)_z$ spin rotation symmetries for uniform s-wave superconductivity. Taken together, our results reveal a rich interplay between quantum geometry and band topology in QGN superconductors.

{\bf Discussions and concluding remarks:}  In  conclusion, our TBCB model provides a minimal realization of a  new class of M-point \mr systems with topological flat bands and momentum-space nonsymmorphic symmetry. This nonsymmorphic symmetry naturally realizes the QGN condition in the PDW channel, and enables us to construct an exact PDW ground state. We further prove a universal relation between the superfluid stiffness and the two-particle effective mass for all QGN superconductors, and derive a topological lower bound for the PDW stiffness in our model. Promising material candidates for our model include twisted $\text{Cu}_2\text{N}$ \cite{doi:10.1021/acs.nanolett.3c01111}, and also the exfoliable Pd$_5$AlI$_2$ \cite{Devarakonda2025}. For Pd$_5$AlI$_2$, integrating out an energetically separated $p_z$ orbital yields an $M$-point QBT when spin-orbit coupling is negligible. Ultracold-atom systems offer another promising route: a twisted bilayer square lattice has recently been realized experimentally using a scheme that is readily adaptable to other lattice geometries \cite{Meng2023}, while theoretical proposals have been developed for implementing checkerboard lattice models in cold-atom platforms \cite{PhysRevLett.132.263402}.

{\bf Acknowledgments:} We thank Jonah Herzog-Arbeitman, Eslam Khalaf and Zhaoyu Han for coordinating the submission of an independent proof establishing the exactness of the superfluid stiffness formula~\cite{han2026}, whose results are consistent with ours where they overlap. This work is supported in part by the NSFC under Grant Nos. 12347107 and 12334003 (Z. W., M.-R. Li, and H. Y.), and the New Cornerstone Science Foundation
through the Xplorer Prize (H. Y.). Z. W. acknowledges support in part from the EPSRC under Grant No. EP/X030881/1.


\bibliography{bib}
\clearpage
\newpage
\widetext
\section {End Matter}
\twocolumngrid
\subsection{Appendix A: Proof of $[\tilde{\eta}^{\dagger},\tilde{S}^{(\mu)}(\mathbf{x})]=0$ }
As discussed in the QGN framework \cite{PhysRevX.14.041004}, a sufficient condition for $[\tilde{\eta}^{\dagger},\tilde{S}^{(\mu)}(\mathbf{x})]=0$ is  $[\hat{\eta}^{\dagger},\hat{S}^{(\mu)}(\mathbf{x})]=0$, where $\hat{\eta}^{\dagger},\hat{S}^{(\mu)}(\mathbf{x})$ are the unprojected operators, as long as the $\hat{\eta}^{\dagger}$ satisfies QGN.  Here we first use an approach different from that in \cite{PhysRevX.14.041004} to prove $[\tilde{\eta}^{\dagger},\tilde{S}^{(\mu)}(\mathbf{x})]=0$ provided the unprojected operators commute: $[\hat{\eta}^{\dagger},\hat{S}^{(\mu)}(\mathbf{x})]=0$ for a QGN order $\hat{\eta}^{\dagger}$. To begin with, the quadratic operators $\tilde{S}^{(\mu)}(\mathbf{x})$ acts only in the flat band space and is related to the projected $\hat{P}\hat{S}^{(\mu)}(\mathbf{x})\hat{P}$ operator as: $\hat{P}\hat{S}^{(\mu)}(\mathbf{x})\hat{P}=\tilde{S}^{(\mu)}(\mathbf{x})+\langle0|\hat{S}^{(\mu)}(\mathbf{x})|0\rangle$. We can directly verify that if $[\eta^{\dagger},\hat{S}^{(\mu)}(\mathbf{x})]=0$, then:
\begin{equation}
\begin{aligned}
     \hat{P}[\eta^{\dagger},\hat{S}^{(\mu)}(\mathbf{x})]\hat{P}&=[\hat{P}\eta^{\dagger}\hat{P},\hat{P}\hat{S}^{(\mu)}(\mathbf{x})\hat{P}]\\
     &=[\tilde{\eta}^{\dagger},\tilde{S}^{(\mu)}(\mathbf{x})+\langle0|\hat{S}^{(\mu)}(\mathbf{x})|0\rangle]\\
     &=[\tilde{\eta}^{\dagger},\tilde{S}^{(\mu)}(\mathbf{x})]=0,
\end{aligned}
\end{equation}
where  we have used the fact that $\hat{P}\hat{S}^{(\mu)}(\mathbf{x})(1-\hat{P})\hat{\eta}^{\dagger}\hat{P}=\hat{P}\hat{\eta}^{\dagger}(1-\hat{P})\hat{S}^{(\mu)}(\mathbf{x})\hat{P}=0$. Indeed, if we write $\hat{P}\hat{S}^{(\mu)}(\mathbf{x})(1-\hat{P})\hat{\eta}^{\dagger}\hat{P}$ explicitly:
\begin{equation}
    \sum_{ijk}|\psi_i\rangle\left(\langle\psi_i|\hat{S}^{(\mu)}(\mathbf{x})|\phi_k\rangle\langle\phi_k|\hat{\eta}^{\dagger}|\psi_j\rangle\right)\langle\psi_j|,
\end{equation}
where $|\psi_i\rangle,|\phi_i\rangle$ belong to the flat band space and the orthogonal space respectively. Nevertheless, there is no state $|\phi_k\rangle\in 1-\hat{P}$ such that both $\langle\psi_i|\hat{S}^{(\mu)}(\mathbf{x})|\phi_k\rangle$ and $\langle\phi_k|\hat{\eta}^{\dagger}|\psi_j\rangle$ are simultaneously nonzero, as $\hat{\eta}^{\dagger}|\psi_j\rangle$ and $\hat{S}^{(\mu)}(\mathbf{x})|\psi_i\rangle$ carry different electron numbers in the upper dispersive bands. Concretely, since $\eta^{\dagger}$ is  block-diagonalized, a nonzero overlap $\langle\phi|\hat{\eta}^{\dagger}|\psi_j\rangle$ requires  $\hat{\eta}^{\dagger}|\psi_j\rangle$ has two electrons in the upper dispersive bands, while $\hat{S}^{(\mu)}(\mathbf{x})|\psi_i\rangle$ has at most one electron in those bands. Similarly, we can prove that $\hat{P}\hat{\eta}^{\dagger}(1-\hat{P})\hat{S}^{(\mu)}(\mathbf{x})\hat{P}= \sum_{ijk}|\psi_i\rangle\left(\langle\psi_i|\hat{\eta}^{\dagger}|\phi_k\rangle\langle\phi_k|\hat{S}^{(\mu)}(\mathbf{x})|\psi_j\rangle\right)\langle\psi_j|=0$.

As a result, in order to prove $[\tilde{\eta}^{\dagger},\tilde{S}^{(\mu)}(\mathbf{x})]=0$ for the PDW order parameter $\tilde{\eta}^{\dagger}$ and  $\tilde{S}^{(\mu)}(\mathbf{x})$ operators we construct, we only need to verify $[\hat{\eta}^{\dagger},\hat{S}^{(\mu)}(\mathbf{x})]=0$. This is equivalent to the matrix equation:
\begin{equation}
 \tau_xs_y\cdot S^{(\mu)}+(S^{(\mu)})^T\cdot \tau_xs_y=0,
 \label{commute_condition}
\end{equation}
where $S^{(\mu)}$ is the matrix form factor of the quadratic operator $\hat{S}^{(\mu)}(\mathbf{x})=\hat{f}^{\dagger}(\mathbf{x})S^{(\mu)}\hat{f}(\mathbf{x})$. It is direct to verify that the condition in Eq. \eqref{commute_condition} is indeed satisfied for all the $\hat{S}^{(\mu)}(\mathbf{x}),\mu=0,1,2..5$ operators in our constructed interaction, and hence $[\tilde{\eta}^{\dagger},\tilde{S}^{(\mu)}(\mathbf{x})]=0$ is proved.

\clearpage
\newpage
\widetext

\begin{bibunit}[apsrev4-2]
\setcounter{equation}{0}
\setcounter{figure}{0}
\setcounter{table}{0}
\renewcommand{\theequation}{S\arabic{equation}}
\renewcommand{\thefigure}{S\arabic{figure}}
\renewcommand{\theHequation}{S\arabic{equation}}
\renewcommand{\theHfigure}{S\arabic{figure}}
\renewcommand{\bibnumfmt}[1]{[S#1]}
\renewcommand{\citenumfont}[1]{S#1}
\makeatletter
\gdef\@extra@b@citeb{@supp}
\gdef\@extra@binfo{@supp}
\makeatother

\begin{center}
{\large\bfseries Supplemental Material for ``Exact pair density wave in topological moire flat bands and universal superfluid stiffness''\par}
\vspace{0.8em}
Zhengzhi Wu$^{1,*}$, Ming-Rui Li$^{2,*}$, and Hong Yao$^{2,\dagger}$\\
\vspace{0.25em}
{\itshape $^1$Rudolf Peierls Centre for Theoretical Physics, Parks Road, Oxford, OX1 3PU, UK\\
$^2$Institute for Advanced Study, Tsinghua University, Beijing 100084, China}\\
\vspace{0.25em}
{\small $^*$These two authors contributed equally to this work.\quad
$^\dagger$\href{mailto:yaohong@tsinghua.edu.cn}{yaohong@tsinghua.edu.cn}}\\
\vspace{0.25em}
(Dated: \today)
\end{center}
\vspace{1.5em}

\onecolumngrid

\section{A. Continuum Model and Symmetry Analysis of the TBCB model}

We consider the twisted bilayer checkerboard lattice (TBCB) model, where the monolayer hosts a quadratic band touching (QBT) described by the single-particle Hamiltonian: $h(\mathbf{k})=t_z(k_x^2-k_y^2)\sigma_z+t_xk_xk_y\sigma_x$. Here, the momentum $\mathbf{k}$ is measured relative to the band-touching point $\frac{1}{a}(\pi,\pi)$ \cite{PhysRevLett.103.046811}, and $\sigma_i$ are Pauli matrices acting on the sublattice degrees of freedom $A,B$. For simplicity, we first analyze the symmetries in the spinless case; the generalization to the spin-$\frac{1}{2}$ case is straightforward. Upon twisting the two stacked layers  by $\frac{l\theta}{2}$ relative to the AA stacking configuration, with $l=\pm$ labeling the top and bottom layers, the non-interacting kinetic energy takes the form:
\begin{equation}
    \hat{H}_0=\sum_{\mathbf{Q}_l,\mathbf{Q}_l^{\prime}}\sum_{\mathbf{k}}\hat{f}_{\mathbf{Q}_l}^{\dagger}(\mathbf{k})h_{\mathbf{Q}_l,\mathbf{Q}_l^{\prime}}(\mathbf{k})\hat{f}_{\mathbf{Q}_l^{\prime}}(\mathbf{k}),
\end{equation}
where $\hat{f}_{\mathbf{Q}_l}^{\dagger}(\mathbf{k}) \equiv \hat{f}_{\mathbf{Q}_l,l,\sigma}^{\dagger}(\mathbf{k})$, with $l=\pm$ denoting the layer index, and satisfies the embedding condition: $\hat{f}_{\mathbf{Q}_l+\mathbf{G}}^{\dagger}(\mathbf{k}+\mathbf{G})=\hat{f}_{\mathbf{Q}_l}^{\dagger}(\mathbf{k})$ with $\mathbf{G}$ the reciprocal momentum of the \mr Brillouin zone. 
The moir\'{e} reciprocal lattice vectors $\mathbf{Q}_l \in \mathcal{Q}_l$ are defined as:
\begin{align}
    \mathcal{Q}_l=\mathbb{Z}\mathbf{b}_{m,1}+\mathbb{Z}\mathbf{b}_{m,2}-\mathbf{M}_{l}.
\end{align}
Here, $\mathbf{M}_l$ specifies the momentum of the band-touching point for layer $l$ with $\mathbf{M}_+=(0,0)$ and $\mathbf{M}_-=-\frac{1}{a_m}(\pi,\pi)$, and the reciprocal momenta  $\mathbf{b}_{m,1}=a_m^{-1}(2\pi,0),\mathbf{b}_{m,2}=a_m^{-1}(0,2\pi)$, where the \mr lattice constant $a_m=\frac{a}{2\sin{\theta/2}}$ and we set $a_m=1$ hereafter. The momenta $\mathcal{Q}_l$ form a square lattice, as illustrated in Fig.~\ref{fig:minibz} (a). Specifically, $\mathcal{Q}_+$ corresponds to the set of $\mathbf{\Gamma}$ points of the mini-Brillouin zone (mBZ), whereas $\mathcal{Q}_-$ corresponds to the $\mathbf{M}=(\pi,\pi)$ points of the mBZ. The matrix elements of the single-particle Hamiltonian, $h_{\mathbf{Q}_l,\mathbf{Q}_l^{\prime}}(\mathbf{k})$, consist of two components:
\begin{equation}
\begin{aligned}
        &h_{\mathbf{Q}_l,\mathbf{Q}_{l'}^{\prime}}(\mathbf{k})=\delta_{\mathbf{Q}_l,\mathbf{Q}_l^{\prime}}h^0_{\mathbf{Q}_l,l}(\mathbf{k})+h_{\mathbf{Q}_l,\mathbf{Q}_{l'}^{\prime}}^{t}(\mathbf{k}),\\
        &h^0_{\mathbf{Q}_l,l}(\mathbf{k})=h(\mathbf{k}-\mathbf{Q}_l),\quad h_{\mathbf{Q}_l,\mathbf{Q}_{l'}^{\prime}}^{t}(\mathbf{k})=\sum_{i=1,2}\delta_{\mathbf{Q}_{l'}^{\prime},\mathbf{Q}_l\pm\mathbf{q}_i}T_i,\\
        &T_1 = \begin{pmatrix} w_{AA} & w_{AB} \\ w_{AB} & w_{AA} \end{pmatrix}, \quad T_2 = \begin{pmatrix} w_{AA} & -w_{AB} \\ -w_{AB} & w_{AA} \end{pmatrix},\quad \mathbf{q}_1=(\pi,\pi),\mathbf{q}_2=(\pi,-\pi)
\end{aligned}
\end{equation}
where $h^0_{\mathbf{Q}_l,l}(\mathbf{k})$ and $h_{\mathbf{Q}_l,\mathbf{Q}_{l'}^{\prime}}^{t}(\mathbf{k})$ denote the intra-layer kinetic energy and the inter-layer hopping, respectively. 

\begin{figure}[htbp]
    \centering
    \includegraphics[width=0.7\linewidth]{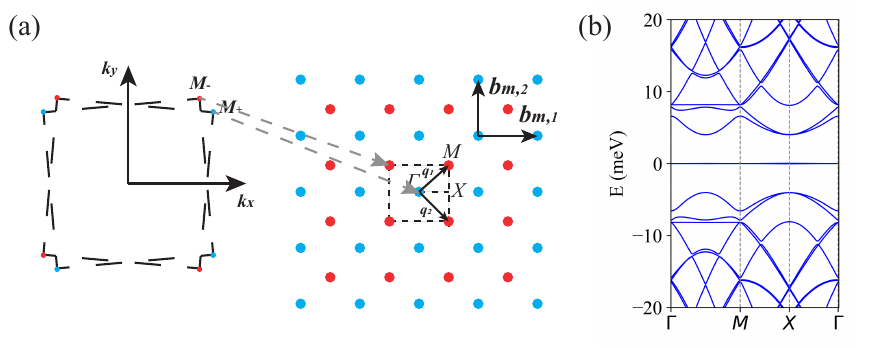}
\caption{(a) Moir\'{e} Brillouin zone of the TBCB model, illustrating the M-point moir\'{e} structure and the finite momentum transfer relevant for the nonsymmorphic symmetry. (b) Band structure in the chiral limit. Two exact flat bands (per spin) lie at zero energy, while the dispersive bands are symmetric with respect to the flat bands owing to the chiral symmetry. }
    \label{fig:minibz}
\end{figure}

In the following, we focus on the chiral limit ($w_{AA}=0$), where exactly flat bands emerge at magic angles. We systematically analyze the relevant symmetries of the system, which are crucial for constructing the exactly solvable models discussed later.

\textbf{(1) Momentum-space nonsymmorphic symmetries $\tilde{C}_{2x/y}$:}
\begin{equation}
    \tilde{C}_{2x} f_{\mathbf{Q}_l,l}(\mathbf{k}) \tilde{C}_{2x}^{-1} = \sigma_z  f_{ C_{2x} \mathbf{Q}_l - \mathbf{q}_{2},-l}(C_{2x} \mathbf{k} - \mathbf{q}_{2}), \quad \tilde{C}_{2y} f_{\mathbf{Q}_l,l}(\mathbf{k}) \tilde{C}_{2y}^{-1} = \sigma_z  f_{C_{2y} \mathbf{Q}_l + \mathbf{q}_{2},-l}(C_{2y} \mathbf{k} + \mathbf{q}_{2}),
\end{equation}
where $C_{2x}\mathbf{k}=(k_x,-k_y)$ and $C_{2y}\mathbf{k}=(-k_x,k_y)$. 

\textbf{(2) Conventional mirror symmetries $M_{x/y}$:}
\begin{equation}
            M_{x/y} f_{\mathbf{Q}_l,l}(\mathbf{k}) M_{x/y}^{-1}= \sigma_{z} f_{M_{x/y} \mathbf{Q}_l,l}(M_{x/y}\mathbf{k}),
\end{equation}
where $M_{x}\mathbf{k}=(-k_x,k_y)$ and $M_{y}\mathbf{k}=(k_x,-k_y)$. 

By combining these operations, we obtain the intrinsic momentum-space nonsymmorphic symmetry $\tilde{M}_z=M_{x/y}\circ \tilde{C}_{2y/x}$, which acts as:
\begin{align}
\tilde{M}_zf_{\mathbf{Q}_l,\sigma,l}(\mathbf{k})\tilde{M}_z^{-1}=f_{\mathbf{Q}_l-\mathbf{q}_1,\sigma,-l}(\mathbf{k}-\mathbf{q}_1).
\end{align}
This symmetry imposes strict constraints on the Bloch wave functions.  The eigen Bloch functions in the single-particle  basis reads:
\begin{equation}
\sum_{\mathbf{Q}^{\prime}_{l'},\sigma^{\prime},l^{\prime}} h_{\mathbf{Q}_{l},\sigma,l; \mathbf{Q}^{\prime}_{l'},\sigma^{\prime},l^{\prime}}(\mathbf{k}) u^{(n)}_{\mathbf{Q}^{\prime}_{l'},\sigma^{\prime},l^{\prime}}(\mathbf{k}) = \epsilon^{(n)}(\mathbf{k}) u^{(n)}_{\mathbf{Q}_l,\sigma,l}(\mathbf{k}),
\end{equation}
where $n$ denotes the band index. The intrinsic momentum-space nonsymmorphic symmetry of $\hat{H}_0$ dictates that the Bloch wavefunctions satisfy:
\begin{equation}
u^{(n)}_{\mathbf{Q}_l,\sigma,l}(\mathbf{k}) = e^{i\theta_{\mathbf{k},n}^M} u^{(n)}_{\mathbf{Q}_l-\mathbf{q}_1,\sigma, -l}(\mathbf{k}-\mathbf{q}_1),
\label{eq:nonsymm_constraint}
\end{equation}
where $\theta_{\mathbf{k},n}^{M}$ is a phase factor arising from the nonsymmorphic translation, depending on both the momentum $\mathbf{k}$ and the band index $n$. This relation explicitly connects the wavefunction amplitude on layer $l$ at momentum $\mathbf{k}$ to that on layer $-l$ at the shifted momentum $\mathbf{k}-\mathbf{q}_1$. In the chiral limit, we employ $n=\pm 1$ to denote the two zero-energy flat bands, while $|n| > 1$ labels the dispersive bands above and below them.

\textbf{(3) Time-reversal symmetry:}
The single-particle Hamiltonian satisfies time-reversal symmetry, $h_{\mathbf{Q}_l,\mathbf{Q}^{\prime}_{l'}}(\mathbf{k})=h_{-\mathbf{Q}_l,-\mathbf{Q}^{\prime}_{l'}}^*(-\mathbf{k})$, which enforces:
\begin{equation}
    u_{\mathbf{Q}_l,\sigma,l}^{(n)}(\mathbf{k})=e^{i\theta^{t}_{\mathbf{k},n}}(u_{-\mathbf{Q}_l,\sigma,l}^{(n)}(-\mathbf{k}))^*
    \label{trs}
\end{equation}

\textbf{(4) Chiral symmetry:}
The Hamiltonian satisfies $\{\sigma_y, h(\mathbf{k})\}=0$ in the chiral limit, yielding the following constraint on the Bloch wavefunctions:
\begin{equation}
    \sigma_y u^{(n)}(\mathbf{k})=e^{i\theta^{c}_{\mathbf{k},n}}u^{(-n)}(\mathbf{k})\quad \text{for }n>0.
    \label{chiral}
\end{equation}

\textbf{(5) $C_4$ rotation symmetry:}
Finally, the model respects a four-fold rotation symmetry $C_{4z}$, acting on the fermions as: 
\begin{align}
C_{4z} f_{\mathbf{Q}_l,l}(\mathbf{k}) C_{4z}^{-1}=\sigma_{y} f_{R_{\pi/2} \mathbf{Q}_l,l}(R_{\pi/2}\mathbf{k}).    
\end{align}

\section{B. Quantum Geometric Nesting in the TBCB Model }

Here we exhaustively enumerate all possible order parameters exhibiting quantum geometric nesting (QGN) in our model, which have the leading ordering tendency \cite{PhysRevX.14.041004}.  Mathematically, an order parameter satisfies perfect QGN if its associated form factor is strictly block-diagonal in the energy band basis, which means there is no off-diagonal form factors which couple the flat band subspace and dispersive band subspace. We find that all possible order parameters with QGN are guaranteed by the symmetries analyzed in the previous section.

Before we dive into the details of the method to find out all possible QGN order parameters, we first remark on some subtleties. At first glance, a plausible order parameter form factor should take the form $N^{\mathbf{Q}_l,\mathbf{Q}_{l^{\prime}}}_{\mu\nu}$, which lives in a rather huge space. Nevertheless, not all the form factors $N^{\mathbf{Q}_l,\mathbf{Q}_{l^{\prime}}}_{\mu\nu}$ give a legitimate order parameter, since a legitimate order parameter, e.g. $\sum_{\mathbf{k},\mathbf{Q}_l,\mathbf{Q}_{l^{\prime}},\mu,\nu}\hat{f}^{\dagger}_{\mathbf{Q}_l,\mu}(\mathbf{k})N^{\mathbf{Q}_l,\mathbf{Q}_{l^{\prime}}}_{\mu\nu}\hat{f}^{\dagger}_{\mathbf{Q}^{\prime}_{l},\nu}(-\mathbf{k})$ should be independent of the embedding, which means it is invariant if we shift the Bloch momentum $\mathbf{k}$ by a reciprocal momentum. As a result, we define all the candidate order parameters in the real space, which circumvents the embedding issue.  We first define the real space fermion operators as:
\begin{equation}
    \hat{f}_{\mu,s}^{\dagger}(\mathbf{x})=\frac{1}{\sqrt{S}}\sum_{\mathbf{k}\in \text{mBZ}}\sum_{\mathbf{Q}_l}\hat{f}_{\mathbf{Q}_l,\mu,s}^{\dagger}(\mathbf{k})e^{-i(\mathbf{k}-\mathbf{Q}_l)\cdot\mathbf{x}}=\frac{1}{\sqrt{S}}\sum_{\mathbf{k}\in \text{mBZ}}\sum_{\mathbf{Q}_l}\sum_n(u^n_{\mathbf{Q}_l,\mu,s}(\mathbf{k}))^*\hat{\gamma}^{\dagger}_{n,s}(\mathbf{k})e^{-i(\mathbf{k}-\mathbf{Q}_l)\cdot\mathbf{x}},
\end{equation}
where $\mathbf{Q}_l=\mathbf{G}-\mathbf{M}_l$ with valley momenta $\mathbf{M}_+=(0,0)$ and $\mathbf{M}_-=-(\pi,\pi)$. Here we use a slightly different definition of the real-space operator $\hat{f}_{\mu}^{\dagger}(\mathbf{x})$ from the maintext to make the QGN order parameter in a simpler form. We put a layer dependent momentum $\mathbf{M}_{l}$ into the  definition of $\hat{f}_{\mu}^{\dagger}(\mathbf{x})$, such that it transforms with a layer-dependent phase under moir\'{e} translations: $\hat{T}_{\mathbf{a}_m}\hat{f}_{\mu}^{\dagger}(\mathbf{x})\hat{T}^{-1}_{\mathbf{a}_m}=\hat{f}_{\mu}^{\dagger}(\mathbf{x}+\mathbf{a}_m)e^{i\mathbf{M}_{l_\mu}\cdot \mathbf{a}_m}$.  We will stick to this definition of $\hat{f}_{\mu}^{\dagger}(\mathbf{x})$ hereafter.  The index $\mu$ encompasses both sublattice and layer degrees of freedom ($\sigma, l$). $u^n_{\mathbf{Q}_l,\mu,s}(\mathbf{k})$ represents the normalized Bloch wavefunction of band $n$, and $\hat{\gamma}^{\dagger}_{n,s}(\mathbf{k})$ is the corresponding band operator.

For a candidate superconducting (particle-particle) order parameter carrying a center-of-mass momentum $\mathbf{q}$, its real-space representation takes the form:
\begin{align}
\eta_{\mathbf{q}}=\sum_{\mathbf{G}}\int d^2\mathbf{x} \hat{f}_{\mu,s}^{\dagger}(\mathbf{x})N^{\mathbf{G}}_{\mu s;\nu s^{\prime}}\hat{f}_{\nu,s^{\prime}}^{\dagger}(\mathbf{x})e^{i(\mathbf{q}+\mathbf{M}_{l_{\mu}}+\mathbf{M}_{l_{\nu}})\cdot \mathbf{x}}e^{i\mathbf{G}\cdot \mathbf{x}},
\end{align}
where $\mathbf{G}\in\mathcal{Q}\equiv\mathbb{Z}\mathbf{b}_{m,1}+\mathbb{Z}\mathbf{b}_{m,2}$. The matrix $N^{\mathbf{G}}_{\mu s; \nu s^{\prime}}$ completely encapsulates the spatial and internal tensor structure of the pairing potential across orbital, layers,  spins and the reciprocal structure $\mathbf{G}$. Written in the energy band basis, it takes:
\begin{equation}
    \begin{aligned}
        \eta_{\mathbf{q}}&=\sum_{\mathbf{k}}\sum_{m,n}\hat{\gamma}^{\dagger}_{m,s}(\frac{\mathbf{q}}{2}+\mathbf{k})F_{ms;ns^{\prime}}(\mathbf{k})\hat{\gamma}^{\dagger}_{n,s^{\prime}}(\frac{\mathbf{q}}{2}-\mathbf{k}),\\
        F_{ms;ns^{\prime}}(\mathbf{k})&=\sum_{\mathbf{G}}\sum_{\mathbf{Q}}(u_{\mathbf{Q}-\mathbf{M}_{l_{\mu}},\mu,s}^m(\frac{\mathbf{q}}{2}+\mathbf{k}))^*N^{\mathbf{G}}_{\mu s,\nu s^{\prime}}(u_{-\mathbf{Q}-\mathbf{G}-\mathbf{M}_{l_{\nu}},\nu,s^{\prime}}^n(\frac{\mathbf{q}}{2}-\mathbf{k}))^*,
    \end{aligned}
\end{equation}
with $\mathbf{Q}\in\mathcal{Q}$. The QGN block-diagonalization condition strictly demands that the cross-band scattering amplitude identically vanishes: $F_{m\in\text{flat},n\notin\text{flat}}(\mathbf{k}) = F_{n\notin\text{flat},m\in\text{flat}}(\mathbf{k}) = 0,\forall \mathbf{k}$. We find  that to find the desired form factor $N^{\mathbf{G}}_{\mu s; \nu s^{\prime}}$ is equivalent to find all the zero modes of the  following positive semi-definite matrix $\Pi^{\text{pp/ph},\mathbf{q}}$:
\begin{equation}
\begin{aligned}
       \Pi_{\mu^{\prime}\nu^{\prime}\mathbf{G}^{\prime},s s^{\prime};\mu\nu\mathbf{G},ss^{\prime}}^{\text{pp}-\mathbf{q}}&=\frac{1}{S}\sum_{\mathbf{k}}\sum_{\mathbf{Q},\mathbf{Q}^{\prime}}\left(\sum_{m\in \text{flat}}u_{\mathbf{Q}^{\prime}-M_{l_\mu^{\prime}},\mu^{\prime},s}^{m}(\frac{\mathbf{q}}{2}+\mathbf{k})(u_{\mathbf{Q}-M_{l_\mu},\mu,s}^{m}(\frac{\mathbf{q}}{2}+\mathbf{k}))^*\right)\\
       &\left(\sum_{n\notin\text{flat}}(u_{-\mathbf{Q}-\mathbf{G}-M_{l_\nu},\nu,s^{\prime}}^{n}(\frac{\mathbf{q}}{2}-\mathbf{k}))^*u_{-\mathbf{Q}^{\prime}-\mathbf{G}^{\prime}-M_{l_\nu^{\prime}},\nu^{\prime},s^{\prime}}^{n}(\frac{\mathbf{q}}{2}-\mathbf{k})\right)\\
       &+\left(\sum_{n\notin\text{flat}}u_{\mathbf{Q}^{\prime}-M_{l_\mu^{\prime}},\mu^{\prime},s}^{n}(\frac{\mathbf{q}}{2}+\mathbf{k})(u_{\mathbf{Q}-M_{l_\mu},\mu,s}^{n}(\frac{\mathbf{q}}{2}+\mathbf{k}))^*\right)\\
       &\left(\sum_{m\in \text{flat}}(u_{-\mathbf{Q}-\mathbf{G}-M_{l_\nu},\nu,s^{\prime}}^{m}(\frac{\mathbf{q}}{2}-\mathbf{k}))^*u_{-\mathbf{Q}^{\prime}-\mathbf{G}^{\prime}-M_{l_\nu^{\prime}},\nu^{\prime},s^{\prime}}^{m}(\frac{\mathbf{q}}{2}-\mathbf{k})\right).
\end{aligned}
\end{equation}
It is straightforward to verify that the matrix $\Pi_{\mu^{\prime}\nu^{\prime}\mathbf{G}^{\prime},s s^{\prime};\mu\nu\mathbf{G},ss^{\prime}}^{\text{pp}-\mathbf{q}}$ is positive semi-definite and hermitian:
\begin{equation}
\begin{aligned}
    \sum_{\mu^{\prime}\nu^{\prime}\mathbf{G}^{\prime};\mu\nu\mathbf{G}}\sum_{s,s^{\prime}} (N_{\mu^{\prime}s;\nu^{\prime}s^{\prime}}^{\mathbf{G}^{\prime}})^*\Pi_{\mu^{\prime}\nu^{\prime}\mathbf{G}^{\prime},s s^{\prime};\mu\nu\mathbf{G},s s^{\prime}}^{\text{pp}-\mathbf{q}}N^{\mathbf{G}}_{\mu s;\nu s^{\prime}}&=\frac{1}{S}\sum_{\mathbf{k}}\sum_{m\in \text{flat};n\notin\text{flat}}|F_{ms;ns^{\prime}}(\mathbf{k})|^2+|F_{ns;ms^{\prime}}(\mathbf{k})|^2\geq0\\
    (\Pi_{\mu\nu\mathbf{G};\mu^{\prime}\nu^{\prime}\mathbf{G}^{\prime}}^{\text{pp}-\mathbf{q}})^*&=\Pi_{\mu^{\prime}\nu^{\prime}\mathbf{G}^{\prime};\mu\nu\mathbf{G}}^{\text{pp}-\mathbf{q}},
\end{aligned}
\end{equation}
where it is clear that a QGN-order parameter is a zero mode of the hermitian matrix $\Pi_{\mu^{\prime}\nu^{\prime}\mathbf{G}^{\prime},s s^{\prime};\mu\nu\mathbf{G},ss^{\prime}}^{\text{pp}-\mathbf{q}}$ from the first line.

A completely analogous formulation applies to the orders in the particle-hole (ph) channel. For a candidate particle-hole order parameter with momentum $\mathbf{q}$, its real-space representation and corresponding band-basis projection take the form:
    \begin{equation}
    \begin{aligned}
        \sum_{\mathbf{G}}\int d^2x \hat{f}_{\mu}^{\dagger}(\mathbf{x})N^{\mathbf{G}}_{\mu\nu}\hat{f}_{\nu}(\mathbf{x})e^{i(\mathbf{q}+\mathbf{M}_{l_{\mu}}-\mathbf{M}_{l_{\nu}})\cdot \mathbf{x}}e^{i\mathbf{G}\cdot 
        \mathbf{x}}&=\sum_{\mathbf{k}}\sum_{m,n}\hat{\gamma}^{\dagger}_{m}(\frac{\mathbf{q}}{2}+\mathbf{k})F_{mn}(\mathbf{k})\hat{\gamma}_{n}(\mathbf{k}-\frac{\mathbf{q}}{2}),\\
        F_{mn}(\mathbf{k})&=\sum_{\mathbf{G}}\sum_{\mathbf{Q}}(u_{\mathbf{Q}-\mathbf{M}_{l_{\mu}},\mu}^m(\frac{\mathbf{q}}{2}+\mathbf{k}))^*N^{\mathbf{G}}_{\mu,\nu}u_{\mathbf{Q}+\mathbf{G}-\mathbf{M}_{l_{\nu}},\nu}^n(\mathbf{k}-\frac{\mathbf{q}}{2}).
    \end{aligned}
\end{equation}
Following the same logic of the superconducting order, we define a positive semi-definite  matrix $\Pi^{\text{ph}-\mathbf{q}}$ for the particle-hole channel order:
\begin{equation}
\begin{aligned}
       \Pi_{\mu^{\prime}\nu^{\prime}\mathbf{G}^{\prime};\mu\nu\mathbf{G}}^{\text{ph}-\mathbf{q}}&=\frac{1}{S}\sum_{\mathbf{k}}\sum_{\mathbf{Q},\mathbf{Q}^{\prime}}\left(\sum_{m\in \text{flat}}u_{\mathbf{Q}^{\prime}-M_{l_\mu^{\prime}},\mu^{\prime}}^{m}(\frac{\mathbf{q}}{2}+\mathbf{k})(u_{\mathbf{Q}-M_{l_\mu},\mu}^{m}(\frac{\mathbf{q}}{2}+\mathbf{k}))^*\right)\\
       &\left(\sum_{n\notin\text{flat}}u_{\mathbf{Q}+\mathbf{G}-M_{l_\nu},\nu}^{n}(\mathbf{k}-\frac{\mathbf{q}}{2}))(u_{\mathbf{Q}^{\prime}+\mathbf{G}^{\prime}-M_{l_\nu^{\prime}},\nu^{\prime}}^{n}(\mathbf{k}-\frac{\mathbf{q}}{2}))^*\right)\\
       +&\left(\sum_{n\notin\text{flat}}u_{\mathbf{Q}^{\prime}-M_{l_\mu^{\prime}},\mu^{\prime}}^{n}(\frac{\mathbf{q}}{2}+\mathbf{k})(u_{\mathbf{Q}-M_{l_\mu},\mu}^{n}(\frac{\mathbf{q}}{2}+\mathbf{k}))^*\right)\\
       &\left(\sum_{m\in \text{flat}}u_{\mathbf{Q}+\mathbf{G}-M_{l_\nu},\nu}^{m}(\mathbf{k}-\frac{\mathbf{q}}{2}))(u_{\mathbf{Q}^{\prime}+\mathbf{G}^{\prime}-M_{l_\nu^{\prime}},\nu^{\prime}}^{m}(\mathbf{k}-\frac{\mathbf{q}}{2}))^*\right)
\end{aligned}
   \end{equation}

Ultimately,  in both the particle-particle and particle-hole channels, to find  QGN order parameters is equivalent to find the zero modes of the following matrices $\Pi^{\text{pp/ph}-\mathbf{q}}$: 
\begin{align}
\Pi^{\text{pp/ph}-\mathbf{q}} N = 0.
\end{align}

Nevertheless, a direct numerical diagonalization of $\Pi^{\text{pp/ph}-\mathbf{q}}$ in the continuum \mr model faces two generic difficulties. First, on a finite $\mathbf{k}$ mesh and with a finite cutoff of the $\mathbf{Q}_l$ space, exact zero modes of $\Pi^{\text{pp/ph}-\mathbf{q}}$ acquire small positive eigenvalues. As a result, we use a small cutoff $\epsilon_{\Pi}$ in the numerics and only retain eigenvalues smaller than $\epsilon_{\Pi}$. Second, because the dispersive-band Hilbert space is much larger than the flat-band sector, the numerical (near)-null space of $\Pi^{\text{pp/ph}-\mathbf{q}}$ contains many trivial solutions whose projected form factors vanish identically in the flat bands. To distill the physically relevant QGN solutions, we use the following two-step numerical procedure

In the pp channel, we first impose on the form factor $N^{\mathbf{G}}_{\mu\nu}$ the constraint dictated by the fermionic anticommutation relations of the superconducting order parameter. Let $\mathcal{P}_{\text{swap}}$ be the permutation operator acting as $N \mapsto N^T$ on the orbital/layer indices. We then diagonalize the penalized matrices
\begin{align}
\tilde{\Pi}^{\text{pp}-\mathbf{q}}_{\mathrm{s/t}}
=
\Pi^{\text{pp}-\mathbf{q}}
+\lambda (I \mp \mathcal{P}_{\text{swap}})^\dagger (I \mp \mathcal{P}_{\text{swap}}),
\end{align}
where the minus (plus) sign corresponds to the singlet (triplet) sector and $\lambda$ is chosen to be a large parameter compared to the numerical near-zero scale $\epsilon_{\Pi}$. The pp (near)-null space is defined as the span of eigenvectors of $\tilde{\Pi}^{\text{pp}-\mathbf{q}}_{\mathrm{s/t}}$ with eigenvalues smaller than a cutoff $\epsilon_{\Pi}$. In the ph channel, no exchange constraint is needed, and we directly define the (near)-null space from $\Pi^{\text{ph}-\mathbf{q}}$ itself.

Secondly, we remove unphysical solutions which has vanishing elements in the flat-band space. For an orthonormal basis $\{N_i\}$ of the (near)-null space, we define the following matrices:
\begin{align}
M^{\text{pp}}_{ij}
=
\frac{1}{N_{\mathbf{k}}}
\sum_{\mathbf{k}}
\Tr\!\left[
\left(P_{\mathbf{k}_+} N_i P_{\mathbf{k}_-}^{T}\right)^\dagger
\left(P_{\mathbf{k}_+} N_j P_{\mathbf{k}_-}^{T}\right)
\right]
\end{align}
for the pp channel, and
\begin{align}
M^{\text{ph}}_{ij}
=
\frac{1}{N_{\mathbf{k}}}
\sum_{\mathbf{k}}
\Tr\!\left[
\left(P_{\mathbf{k}_+} N_i P_{\mathbf{k}_-}\right)^\dagger
\left(P_{\mathbf{k}_+} N_j P_{\mathbf{k}_-}\right)
\right]
\end{align}
for the ph channel. The parameter $N_{\mathbf{k}}$ is the number of momentum points sampled in the \mr Brillouin zone in numerics. Here $P_{\mathbf{k}_+}$ and $P_{\mathbf{k}_-}$ denote the flat-band projectors evaluated at $\mathbf{k}_+=\mathbf{q}/2+\mathbf{k}$ and $\mathbf{k}_-=\mathbf{q}/2-\mathbf{k}$ in the pp channel, while in the ph channel $\mathbf{k}_-$ is understood as $\mathbf{k}-\mathbf{q}/2$. The positive eigenvalues of $M$ measure the flat-band spectral weight of orthogonal linear combinations inside the numerical near-null space. We therefore retain all eigenvectors of $M$ with eigenvalues larger than a small threshold $\epsilon_M$. This step removes unphysical solutions with zero elements in the flat-band space, while preserving the QGN block diagonalized structure.

Applying this procedure to the TBCB model, we find that all possible QGN order parameters carry momentum either $\mathbf{q}=\mathbf{\Gamma}=0$ or $\mathbf{q}=\mathbf{M}=(\pi,\pi)$. We start with the QGN orders in the pp channel. For $\mathbf{q}=0$, the pp channel yields one singlet and one triplet uniform pairings, which can be chosen as:
\begin{align}
N_{\mathbf{\Gamma}}^{\text{singlet}}=\tau_0 \sigma_0s_y,
\qquad
N_{\mathbf{\Gamma}}^{\text{triplet}}= i\tau_0 \sigma_ys_y\vec{s}.
\end{align}
 Since both orders are diagonal in the layer space, they describe uniform intra-layer pairings: an intra-sublattice singlet pairing $N_{\mathbf{\Gamma}}^{\text{singlet}}$ and an inter-sublattice triplet pairing $N_{\mathbf{\Gamma}}^{\text{triplet}}$.

At finite momentum $\mathbf{q}=\mathbf{M}$, the pp channel yields two PDW orders:
\begin{align}
N_{\mathbf{M}}^{\text{singlet}}=\tau_x \sigma_0s_y,
\qquad
N_{\mathbf{M}}^{\text{triplet}}= i\tau_x \sigma_ys_y\vec{s}.
\end{align}
These PDW orders are inter-layer pairings, which are enforced by the momentum-space nonsymmorphic symmetry. In particular, the singlet PDW $N_{\mathbf{M}}^{\text{singlet}}=\tau_x \sigma_0s_y$ is exactly the inter-layer PDW operator $\eta^\dagger$ constructed in the main text. With an appropriate parent interaction, one can also construct an exact triplet PDW ground state corresponding to $N_{\mathbf{M}}^{\text{triplet}}$. We leave this construction to future work.

In the ph channel,  QGN order parameters also carry momentum $\mathbf{q}=\mathbf{\Gamma}=0$ or $\mathbf{q}=\mathbf{M}=(\pi,\pi)$. At $\mathbf{q}=0$, there are two classes of orders:
\begin{align}
N_{\mathbf{\Gamma},1}^{\text{ph}}=\tau_0 \sigma_0s^\mu,
\qquad
N_{\mathbf{\Gamma},2}^{\text{ph}}= \tau_0 \sigma_ys^\mu,\quad\mu=0,x,y,z
\end{align}
The first class of orders are uniform spin orders (when $\mu=x,y,z$), whereas the second class of orders correspond to QAH and QSH orders.  At $\mathbf{q}=\mathbf{M}$, we obtain two density wave orders:
\begin{align}
N_{\mathbf{M},1}^{\text{ph}}=\tau_x \sigma_0s^\mu,
\qquad
N_{\mathbf{M},2}^{\text{ph}}= \tau_x \sigma_ys^\mu,
\end{align}
which describe inter-layer density waves carrying momentum $(\pi,\pi)$.

The QGN nature of the above order parameters is guaranteed by the symmetries analyzed in the previous section. In particular, using the three symmetries in Eq.~\eqref{eq:nonsymm_constraint}, Eq.~\eqref{trs}, and Eq.~\eqref{chiral}, we can  analytically verify  that the corresponding band-basis form factors are block diagonal between the flat-band and remote-band sectors.

\section{C. Detailed Analysis of the Exact  PDW Ground State}

As mentioned in the main text, the PDW ground state is the exact ground state of the projected Hamiltonian $\hat{P}\hat{H}^{\text{int}}\hat{P}$, where 
\begin{equation}
    \begin{aligned}
\hat{{H}}^{int}=U\int d^2x \left[(S^{(0)}(x))^2+\frac{1}{3}\sum_i(S^{(i)}(x))^2\right],\label{eq:interaction_hamiltonian_pdw} \quad  S^{(0)}(x)=f^{\dagger}(x)\tau_zf(x),\quad S^{(i)}(x)=f^{\dagger}(x)\tau_z\sigma_ys^{i}f(x),
\end{aligned}
\end{equation}
provided that  no quadratic one-body terms arise from the projection.
In this section, we rigorously prove the absence of this kind of quadratic one-body terms, and discuss about whether there are other degenerate ground states. 

\subsection{Rigorous Vanishing of Projected One-Body Terms}
\label{sec:sm_vanishing_projected_one_body}
Upon projection onto the flat-band subspace, a four-fermion interaction such as $\hat{V}^{int}=\int d^2x (S^{(\mu)}(x))^2,\mu=0,x,y,z$ with $S^{(\mu)}(x)$ the quadratic operator in Eq. \eqref{eq:interaction_hamiltonian_pdw}, generally does not reduce to $\int d^2x (\tilde{S}^{(\mu)}(x))^2$, with $\tilde{S}^{(\mu)}(x)$ the projected operator only acting in the flat-band space; rather, it also generates additional Hartree-Fock type one-body terms \cite{PhysRevX.14.041004}. Concretely, the general form of a projected four-fermion takes:
\begin{equation}
    \hat{P}\left(\hat{f}_{1}^{\dagger} \hat{f}_{2} \hat{f}_{3}^{\dagger} \hat{f}_{4}\right)\hat{P} = \tilde{f}_{1}^{\dagger} \tilde{f}_{2} \tilde{f}_{3}^{\dagger} \tilde{f}_{4} + \underbrace{\langle\bar{f}_{1}^{\dagger} \bar{f}_{2}\rangle \tilde{f}_{3}^{\dagger} \tilde{f}_{4} + \langle\bar{f}_{3}^{\dagger} \bar{f}_{4}\rangle \tilde{f}_{1}^{\dagger} \tilde{f}_{2}}_{\text{Trace terms}} + \underbrace{\langle\bar{f}_{2} \bar{f}_{3}^{\dagger}\rangle \tilde{f}_{1}^{\dagger} \tilde{f}_{4} + \langle\bar{f}_{1}^{\dagger} \bar{f}_{4}\rangle \tilde{f}_{2} \tilde{f}_{3}^{\dagger}}_{\text{Cross terms}} + \text{const},
\end{equation}
where $\tilde{f}_{i}^{\dagger},\bar{f}_i$ represents the fermion operator acting in the flat-band and dispersive-band subspace respectively: $\tilde{f}_i \equiv \hat{P}\hat{f}_i,\bar{f}_i \equiv (1-\hat{P})\hat{f}_i$, and $\hat{f}_i=\tilde{f}_i+\bar{f}_i$. The expectation value is evaluated on the fully-occupied (empty) dispersive bands whose energies lie below (above) than the flat-band energy. The 'trace terms' and 'cross terms' are the quadratic terms we  focus on in this section.

For our model, the interaction is $\hat{H}^{\text{int}} = U \int d^2\mathbf{x} \left[ (S^{(0)}(\mathbf{x}))^2 + \frac{1}{3}\sum_{i=1}^3 (S^{(i)}(\mathbf{x}))^2 \right]$. 
We first consider the `cross terms' which originate from contractions between different $\hat{S}^{(\mu)}$ operators. First, the contraction $\langle \bar{f}(\mathbf{k})\bar{f}^{\dagger}(\mathbf{k}^{\prime})\rangle$ is non-zero only when evaluated on the empty upper dispersive bands ($n_+$), yielding the upper dispersive-band projection matrix $Q^{+}_{\mathbf{Q},\mathbf{Q}^{\prime}}(\mathbf{k}) = \sum_{n_+} u^{n_+}_{\mathbf{Q}}(\mathbf{k})(u^{n_+}_{\mathbf{Q}^{\prime}}(\mathbf{k}))^*$, where for simplicity, we omit the layer, sublattice and spin indices $l,\sigma,s$ and they are implicitly summed over in the following derivations when we take matrix/operator multiplications. The corresponding one-body term acting on the flat bands takes the form:
\begin{equation}
    \mathcal{T}_{1} = \sum_{\mathbf{k},\mathbf{q},\mathbf{G}} \tilde{f}_{\mathbf{Q}+\mathbf{G}}^{\dagger}(\mathbf{k}+\mathbf{q}) \left[ \tau_z Q^{+}_{\mathbf{Q},\mathbf{Q}^{\prime}}(\mathbf{k}) \tau_z + \frac{1}{3}\sum_{i=1}^3 (\tau_z\sigma_y s^i) Q^{+}_{\mathbf{Q},\mathbf{Q}^{\prime}}(\mathbf{k}) (\tau_z\sigma_y s^i) \right] \tilde{f}_{\mathbf{Q}^{\prime}+\mathbf{G}}(\mathbf{k}+\mathbf{q}).
    \label{eq:term1}
\end{equation}
Due to the global SU(2) spin rotation symmetry, the spin summation simplifies exactly as $\frac{1}{3}\sum_i s^i Q^{+} s^i = Q^{+}$. Thus, the matrix in the bracket reduces to $\tau_z Q^{+} \tau_z + \tau_z\sigma_y Q^{+} \tau_z\sigma_y$.

Meanwhile, the other 'cross term' involves $\langle \bar{f}^{\dagger}(\mathbf{k}^{\prime})\bar{f}(\mathbf{k})\rangle$, which is evaluated on the fully occupied lower dispersive bands ($n_-$). Crucially, putting the remaining flat-band operators $\tilde{f}^{\dagger}$ and $\tilde{f}$ into the normal order introduces a relative minus sign due to fermionic anti-commutation relation. Evaluating this contribution gives:
\begin{equation}
    \mathcal{T}_{2} = - \sum_{\mathbf{k},\mathbf{q},\mathbf{G}} \tilde{f}_{\mathbf{Q}+\mathbf{G}}^{\dagger}(\mathbf{k}+\mathbf{q}) \left[ \tau_z Q^{-}_{\mathbf{Q},\mathbf{Q}^{\prime}}(\mathbf{k}) \tau_z + \tau_z\sigma_y Q^{-}_{\mathbf{Q},\mathbf{Q}^{\prime}}(\mathbf{k}) \tau_z\sigma_y \right] \tilde{f}_{\mathbf{Q}^{\prime}+\mathbf{G}}(\mathbf{k}+\mathbf{q}).
    \label{eq:term2}
\end{equation}

Note that the chiral symmetry rigorously enforces $\sigma_y Q^{+}_{\mathbf{Q},\mathbf{Q}^{\prime}}(\mathbf{k})\sigma_y = Q^{-}_{\mathbf{Q},\mathbf{Q}^{\prime}}(\mathbf{k})$, which maps the upper dispersive-band projector exactly to the lower dispersive-band projector. As a result, we can rewrite the matrix appearing in the bracket of $\mathcal{T}_1$ as:
\begin{equation}
    \tau_z Q^{+} \tau_z + \tau_z\sigma_y Q^{+} \tau_z\sigma_y = \tau_z\sigma_y Q^{-} \tau_z\sigma_y + \tau_z Q^{-} \tau_z.
\end{equation}
Now it is clear that $\mathcal{T}_1$ cancels with  $\mathcal{T}_2$: $\mathcal{T}_1 + \mathcal{T}_2 = 0$. The interactions from the density channel ($S^{(0)}$) and the spin channels ($S^{(i)}$) perfectly conspire with the chiral symmetry to exactly cancel the cross terms.

The remaining quadratic contributions arise from the trace contractions, where $\langle \bar{f}^{\dagger} \bar{f} \rangle$ is contracted within the same $S^{(\mu)}$ operator. Because the upper bands are empty, this expectation value only accumulates over the fully filled lower dispersive bands ($n_-$):
\begin{equation}
    \mathcal{T}_{3} = 2\sum_{\mathbf{G},\mu}a_{\mu}\left(\sum_{\mathbf{Q},\mathbf{k},n_-} (u_{\mathbf{Q}+\mathbf{G}}^{n-}(\mathbf{k}))^{\dagger} S^{(\mu)} u_{\mathbf{Q}}^{n-}(\mathbf{k})\right)\left(\sum_{\mathbf{k}^{\prime}}\tilde{f}^{\dagger}_{\mathbf{Q}^{\prime}}(\mathbf{k}^{\prime})S^{(\mu)}\tilde{f}_{\mathbf{Q}^{\prime}+\mathbf{G}}(\mathbf{k}^{\prime})\right),
\end{equation}
where $\mu=0,x,y,z$ and the coefficients $a_0=1,a_x=a_y=a_z=\frac{1}{3}$. Crucially, all four matrices $S^{(\mu)} \in \{\tau_z\sigma_0s^0, \tau_z\sigma_ys^i\}$ contain $\tau_z$. As established in Eq.~\ref{eq:nonsymm_constraint}, the system preserves a momentum-space nonsymmorphic symmetry $\tilde{M}_z$ which effectively exchanges the two layers ($\tau_z \to -\tau_z$). As a result,  this symmetry rigorously guarantees the coefficient $\left(\sum_{\mathbf{Q},\mathbf{k},n_-} (u_{\mathbf{Q}+\mathbf{G}}^{n-}(\mathbf{k}))^{\dagger} S^{(\mu)} u_{\mathbf{Q}}^{n-}(\mathbf{k})\right)$ is zero  for any reciprocal vector $\mathbf{G}$. 

Consequently, $\mathcal{T}_3 = 0$. Since the additional one-body terms generated by the remote bands cancel, the projected Hamiltonian $\hat{P}\hat{H}^{\text{int}}\hat{P}$ reduces, up to a constant, to the sum of squared projected bilinears within the flat-band subspace. Together with the conclusion $[\tilde{S}^{(\mu)}(x),\tilde{\eta}^{\dagger}]=0$ proved in the main text, we complete the  proof that the PDW ground state is an exact ground state in the strong-coupling flat-band limit.

\subsection{Uniqueness of the PDW ground state}
Here we demonstrate that the exact PDW ground state is the unique ground state of our model $\hat{P}\hat{H}_{\text{int}}\hat{P}$. We restrict our attention to candidate ordered states whose order parameters satisfy the QGN condition, as these constitute the leading ordering tendencies, with susceptibilities saturating the theoretical upper bound \cite{PhysRevX.14.041004}.

We have listed all possible QGN order parameters in the previous section. First, in the pp-channel, there are no other superconducting ground states except the PDW ground state we construct, since only the PDW order parameter commutes with all the $\hat{S}^{(\mu)}(x)$.  As a result, we only focus on the ordered states in the ph-channel.

We first give a brief review of the QGN framework in the ph-channel \cite{PhysRevX.14.041004}, which is closely analogous to that in the pp channel. We still consider a positive semi-definite interaction and try to solve the effective Hamiltonian which is the projected interaction in the flat band space, such as $\hat{H}_{\text{eff}}=\int d^2x (\tilde{S}^{(\mu)}(x))^2$ with $\tilde{S}^{(\mu)}(x)$ the projected quadratic operator only containing fermionic operators $\tilde{f},\tilde{f}^{\dagger}$ acting in the flat band space. We then consider a Hermitian  order parameter in the ph-channel satisfying the QGN condition (Hermiticity is indeed satisfied for all the QGN ph-orders in our TBCB model), such as  $\hat{O}=\int d^2x \hat{f}^{\dagger}(x)N^{p-h}\hat{f}(x)$ with $N^{p-h}$ a matrix in the sublattice, layer and spin space in our model, and still require the flat-band projected order parameter $\tilde{O}=\int d^2x \tilde{f}^{\dagger}(x)N^{p-h}\tilde{f}(x)$ commutes with all the $\tilde{S}^{(\mu)}(x)$. A sufficient condition is that the corresponding unprojected operators satisfy: $[\hat{S}^{(\mu)}(x),\hat{O}]=0$, since $\hat{O}$ satisfies QGN. Consequently, any nondegenerate eigenstate $|\psi\rangle$ of $\tilde{O}$ is also an eigenstate of all the  $\tilde{S}^{(\mu)}(x)$, and $|\psi\rangle$ of $\tilde{O}$ will become a ground state of $\hat{H}_{\text{eff}}$ if it is actually the zero mode of all the $\tilde{S}^{(\mu)}(x)$.


Now we return to our TBCB model and focus on the effective Hamiltonian $\hat{H}_{\text{eff}}=\hat{P}\hat{H}_{\text{int}}\hat{P}$, with $\hat{H}_{\text{int}}$ given  in Eq. \eqref{eq:interaction_hamiltonian_pdw}. Each $\hat{S}^{(\mu)}(x)$ operator takes the form: $\hat{S}^{(\mu)}(x)=\hat{f}^{\dagger}(x)S^{(\mu)}\hat{f}(x)$, and the condition $[\hat{S}^{(\mu)}(x),\hat{O}]=0$ is equivalent to :
\begin{align}
    N^{p-h} \cdot S^{(\mu)} -S^{(\mu)} \cdot N^{p-h}= 0.
\end{align}
Among all the QGN order parameters in the ph-channel, only the QAH order parameter $N_{\text{QAH}} = \tau_0\sigma_y$ satisfies the above condition. The QAH order parameter $\tilde{O}_{\text{QAH}}=\int d^2x \tilde{f}^{\dagger}(x)N_{\text{QAH}}\tilde{f}(x)$ has two nondegenerate eigenstates $|\psi_{\pm}\rangle$, when the flat bands are half-filled and all the positive (negative) eigenstates are fully filled. $|\psi_{\pm}\rangle$ are two QAH states related by the time-reversal symmetry, which are also zero modes of the effective Hamiltonian $\hat{H}_{\text{eff}}=\hat{P}\hat{H}_{\text{int}}\hat{P}$.  



We  can lift this accidental degeneracy between the QAH states and the PDW states at half-filling, and  uniquely select the PDW state as the ground state across all fillings by introducing an infinitesimally small perturbation:
\begin{equation}
    \Delta\hat{\mathcal{H}}=\Delta U\int d^2\mathbf{x} \left[ (S^{(4)}(\mathbf{x}))^2+(S^{(5)}(\mathbf{x}))^2 \right],\quad 0<\Delta U\ll U
\end{equation}
where $S^{(4)}(\mathbf{x})=\hat{f}^{\dagger}(\mathbf{x})\tau_z\sigma_x\hat{f}(\mathbf{x})$ and $S^{(5)}(\mathbf{x})=\hat{f}^{\dagger}(\mathbf{x})\tau_z\sigma_z\hat{f}(\mathbf{x})$. This specific perturbation energetically penalizes the QAH state (since $\sigma_y$ anti-commutes with $\sigma_x$ and $\sigma_z$, breaking the zero-energy condition), while still favors  the PDW state as the exact ground state (zero mode), which guarantees the selection of a unique PDW ground state.

\section{D. Exact Excitation Spectra}
The few-body excitations are also exactly solvable in our model $\hat{H}_{\text{eff}}=\hat{P}\hat{H}^{int}\hat{P}$ with $\hat{H}^{int}$ in Eq. \eqref{eq:interaction_hamiltonian_pdw}. The  eigen-equation of a few-body excitation with charge $c$ and momentum $\mathbf{p}$ is:
\begin{align}
[\hat{{H}}_{\text{eff}},\hat{\mathcal{C}}^c_{\mathbf{p}}]\ket{\Psi}=E_{c}(\mathbf{p})\hat{\mathcal{C}}^c_{\mathbf{p}}\ket{\Psi},
\end{align}
where $\ket{\Psi}$ is the exact ground state and $\hat{\mathcal{C}}^c_{\mathbf{p}}$ is an eigen-excitation operator .

\subsection{Single particle/hole excitation}

 We first analyze the single-particle sector. The effective Hamiltonian can be rewritten in the band basis as:
\begin{align}
 \hat{H}_{\text{eff}} &= \frac{U}{S} \sum_{\mathbf{q} \in mBZ} \sum_{\mathbf{G}} \sum_{\mu=0}^3 a_\mu \hat{\mathcal{S}}^{(\mu)}(\mathbf{q}+\mathbf{G}) \hat{\mathcal{S}}^{(\mu)}(-\mathbf{q}-\mathbf{G}),\\
\hat{\mathcal{S}}^{(\mu)}(\mathbf{q}+\mathbf{G}) &= \sum_{\mathbf{k}, s, s'} \sum_{m, n\in \text{flat}} \Lambda_{m,s; n,s'}^{(\mu)}(\mathbf{k}, \mathbf{q}+\mathbf{G}) \gamma_{m, s}^\dagger(\mathbf{k}+\mathbf{q}) \gamma_{n, s'}(\mathbf{k}),\\
\Lambda_{m,s; n,s'}^{(\mu)}(\mathbf{k}, \mathbf{q}+\mathbf{G}) &= \sum_{\mathbf{Q}} \sum_{l, \sigma, \sigma'} \Big[ \left( u_{\mathbf{Q}-\mathbf{G}, l, \sigma, s}^{(m)}(\mathbf{k}+\mathbf{q}) \right)^*  \left( S^{(\mu)} \right)_{l\sigma s; l\sigma' s'} u_{\mathbf{Q}, l, \sigma', s'}^{(n)}(\mathbf{k}) \Big].
\end{align}
where $a_0=1$  and $a_{1,2,3}=1/3$ . To find single-hole excitations with momentum $\mathbf{P}$, we seek an eigen-operator in the space of $\hat{\gamma}_{m,s}(\mathbf{P})$. Using the  property that the ground state is annihilated by $\hat{\mathcal{S}}^{(\mu)}(\mathbf{q})$:  $\hat{\mathcal{S}}^{(\mu)}(\mathbf{q})|\Psi\rangle=0$, the commutator equation is simplified to:
\begin{align}
    [\hat{H}_{\text{eff}}, \hat{\gamma}_{m,s}(\mathbf{P})]|\Psi\rangle = \frac{U}{S} \sum_{\mathbf{q}, \mathbf{G}, \mu} a_\mu \hat{\mathcal{S}}^{(\mu)}(\mathbf{q}+\mathbf{G}) [\hat{\mathcal{S}}^{(\mu)}(-\mathbf{q}-\mathbf{G}), \hat{\gamma}_{m,s}(\mathbf{P})] |\Psi\rangle.
\end{align}
Evaluating the above commutator yields a linear combination of hole operators shifted by the momentum $\mathbf{q}$:
\begin{align}
    [\hat{\mathcal{S}}^{(\mu)}(-\mathbf{q}-\mathbf{G}), \hat{\gamma}_{m,s}(\mathbf{P})] = -\sum_{n\in\text{flat}, s'} \Lambda^{(\mu)}_{m,s;n,s'}(\mathbf{P}+\mathbf{q}, -\mathbf{q}-\mathbf{G}) \hat{\gamma}_{n,s'}(\mathbf{P}+\mathbf{q}).
\end{align}
This is the first step of the closure. Although the intermediate hole carries momentum $\mathbf{P}+\mathbf{q}$, the other density operator in $\hat{H}_{\text{eff}}$ carries the opposite momentum transfer $\mathbf{q}+\mathbf{G}$. Since the exact ground state is a zero mode,
\begin{align}
    \hat{\mathcal{S}}^{(\mu)}(\mathbf{q}+\mathbf{G})|\Psi\rangle=0,
\end{align}
we can move the remaining $\hat{\mathcal{S}}^{(\mu)}(\mathbf{q}+\mathbf{G})$ through the intermediate hole operator:
\begin{align}
    \hat{\mathcal{S}}^{(\mu)}(\mathbf{q}+\mathbf{G})
    \hat{\gamma}_{n,s'}(\mathbf{P}+\mathbf{q})|\Psi\rangle
    =
    [\hat{\mathcal{S}}^{(\mu)}(\mathbf{q}+\mathbf{G}),
    \hat{\gamma}_{n,s'}(\mathbf{P}+\mathbf{q})]|\Psi\rangle .
\end{align}
The second commutator removes the same momentum $\mathbf{q}$ and therefore returns the operator to the original total momentum $\mathbf{P}$:
\begin{align}
    [\hat{\mathcal{S}}^{(\mu)}(\mathbf{q}+\mathbf{G}), \hat{\gamma}_{n,s'}(\mathbf{P}+\mathbf{q})]
    = -\sum_{l\in\text{flat}, s''} \Lambda^{(\mu)}_{n,s';l,s''}(\mathbf{P}, \mathbf{q}+\mathbf{G}) \hat{\gamma}_{l,s''}(\mathbf{P}).
\end{align}
Therefore the action of the commutator never leaves the single-hole operator space. It only mixes the flat-band labels and spin indices at fixed momentum $\mathbf{P}$, which is why the equation of motion is closed and can be written as a finite scattering matrix. We can then arrive at the following eigen-equation for single hole excitation:
\begin{align}
    [\hat{H}_{\text{eff}}, \hat{\gamma}_{m,s}(\mathbf{P})]|\Psi\rangle &= \sum_{n\in\text{flat},s'} \Gamma_{m,s;n,s'}^{h,\mathbf{P}} \hat{\gamma}_{n,s'}(\mathbf{P})|\Psi\rangle,\\
    \Gamma_{m,s;n,s'}^{h,\mathbf{P}} &= \frac{U}{S} \sum_{\mu=0}^3 a_\mu \sum_{\mathbf{q}, \mathbf{G}} \sum_{l \in \text{flat}, s''} \Lambda^{(\mu)}_{m,s;l,s''}(\mathbf{P}+\mathbf{q}, -\mathbf{q}-\mathbf{G}) \Lambda^{(\mu)}_{l,s'';n,s'}(\mathbf{P}, \mathbf{q}+\mathbf{G}).
    \label{sm_single_hole}
\end{align}

Similarly, we can derive the eigen-equation for the single-electron excitation with momentum $\mathbf{P}$:
\begin{align}
        [\hat{H}_{\text{eff}}, \hat{\gamma}_{m,s}^\dagger(\mathbf{P})]|\Psi\rangle &= \sum_{n\in\text{flat},s'} \Gamma_{m,s;n,s'}^{p,\mathbf{P}} \hat{\gamma}^\dagger_{n,s'}(\mathbf{P})|\Psi\rangle,\\
            \Gamma_{m,s;n,s'}^{p,\mathbf{P}} &= \frac{U}{S} \sum_{\mu=0}^3 a_\mu \sum_{\mathbf{q}, \mathbf{G}} \sum_{l \in \text{flat}, s''} \Lambda^{(\mu)}_{n,s';l,s''}(\mathbf{P}-\mathbf{q}, \mathbf{q}+\mathbf{G}) \Lambda^{(\mu)}_{l,s'';m,s}(\mathbf{P}, -\mathbf{q}-\mathbf{G}),
            \label{sm_single_particle}
\end{align}
By diagonalizing the two matrices $\Gamma_{m,s;n,s'}^{h,\mathbf{P}},\Gamma_{m,s;n,s'}^{p,\mathbf{P}}$ in Eq. \eqref{sm_single_hole}  and \eqref{sm_single_particle}, we can obtain the exact single-particle excitations. Crucially, the sum over the reciprocal momentum $\mathbf{G}$  makes the single-particle excitation dispersive, in contrast to the flat single-particle dispersion in finite-orbital QGN models \cite{PhysRevX.14.041004}.

\subsection{Two-Particle Collective Excitations}
Having obtained the single-particle excitation energy, we can also exactly obtain the two-particle excitation energy, including both charge-neutral particle-hole excitations (excitons) and charge-$2e$ particle-particle excitations (Cooper pairs).
In this section, we first write the spin-resolved equations and then project each channel to spin-0 and spin-1 sectors.  

We begin with the charge-0 particle-hole excitations. The basis of excitons are: $\hat{\Phi}^\dagger_{\mathbf{P}}(\mathbf{k}) = \hat{\gamma}^\dagger_{n,s}(\mathbf{k}+\mathbf{P}) \hat{\gamma}_{m,s'}(\mathbf{k})$, generating an electron-hole pair with center-of-mass momentum $\mathbf{P}$. The corresponding eigen-equation is:
\begin{align}
    [\hat{H}_{\text{eff}}, \hat{\Phi}^\dagger_{\mathbf{P}}(\mathbf{k})]|\Psi\rangle &= \sum_{\mathbf{k}', n'm'\in\text{flat}, s''s'''} \Gamma^{p-h, \mathbf{P}}_{ns, ms'; n's'', m's'''}(\mathbf{k}, \mathbf{k}')\hat{\Phi}^\dagger_{\mathbf{P}}(\mathbf{k}')|\Psi\rangle,\\
    \Gamma^{p-h, \mathbf{P}}_{ns, ms'; n's'', m's'''}(\mathbf{k}, \mathbf{k}') &= \delta_{\mathbf{k}\mathbf{k}'} \delta_{mm'} \delta_{s's'''} \Gamma^{p, \mathbf{k}+\mathbf{P}}_{ns; n's''} + \delta_{\mathbf{k}\mathbf{k}'} \delta_{nn'} \delta_{ss''} \Gamma^{h, \mathbf{k}}_{ms';m's'''}  + \mathcal{V}^{p-h}_{\mathbf{P}}(\mathbf{k}, \mathbf{k}')_{ns, ms'; n's'', m's'''},
    \label{sm_ph_excitation}
\end{align}
where $\Gamma^{p/h, \mathbf{k}}_{ns; n's''}$ are the matrices in Eq. \eqref{sm_single_hole} and \eqref{sm_single_particle}. And the momentum off-diagonal matrix $\mathcal{V}^{p-h}$ takes:
\begin{align}
    \mathcal{V}^{p-h}_{\mathbf{P}}(\mathbf{k}, \mathbf{k}') = -\frac{2U}{S} \sum_{\mu=0}^3  a_\mu \sum_{\mathbf{G}} \Lambda^{(\mu)}_{n',s''; n,s}(\mathbf{k}+\mathbf{P}, -\mathbf{q}-\mathbf{G}) \Lambda^{(\mu)}_{m,s'; m',s'''}(\mathbf{k}', \mathbf{q}+\mathbf{G}).
\end{align}
Here, $\mathbf{q} = \mathbf{k} - \mathbf{k}'$ represents the momentum transfer. 
The prefactor of two follows from the two equivalent commutators with the two density operators in $\hat{\mathcal{S}}^{(\mu)}(\mathbf{q}+\mathbf{G})\hat{\mathcal{S}}^{(\mu)}(-\mathbf{q}-\mathbf{G})$.

The spin-0 and spin-1 particle-hole spectra are obtained by projecting the spin structure onto $\delta_{ss'}$ and $s^a_{ss'}$, respectively.  Since the Bloch wave functions are spin independent, we write
\begin{align}
    \Lambda^{(0)}_{m,s;n,s'}(\mathbf{k},\mathbf{Q})
    =\lambda^{d}_{mn}(\mathbf{k},\mathbf{Q})\delta_{ss'},\quad
    \Lambda^{(i)}_{m,s;n,s'}(\mathbf{k},\mathbf{Q})
    =\lambda^{s}_{mn}(\mathbf{k},\mathbf{Q})(s^i)_{ss'},\quad i=x,y,z,
    \label{sm_lambda_spin_factorization}
\end{align}
where $\lambda^{d}$ and $\lambda^{s}$ are the orbital-layer form factors associated with $\tau_z\sigma_0$ and $\tau_z\sigma_y$.  Using the SU(2) identity
\begin{align}
    \sum_{i=x,y,z}(s^i)_{\alpha\beta}(s^i)_{\gamma\delta}
    =
    2\delta_{\alpha\delta}\delta_{\beta\gamma}
    -\delta_{\alpha\beta}\delta_{\gamma\delta},
\end{align}
we obtain
\begin{align}
    \Gamma^{p-h,\alpha}_{nm;n'm'}(\mathbf{k},\mathbf{k}')
    &=
    \delta_{\mathbf{k}\mathbf{k}'}\delta_{mm'}\Gamma^{p,\mathbf{k}+\mathbf{P}}_{nn'}
    +
    \delta_{\mathbf{k}\mathbf{k}'}\delta_{nn'}\Gamma^{h,\mathbf{k}}_{mm'}
    -\frac{2U}{S}\left[
    D^{p-h}_{\mathbf{P}}(\mathbf{k},\mathbf{k}')_{nm;n'm'}
    +c^{p-h}_{\alpha}X^{p-h}_{\mathbf{P}}(\mathbf{k},\mathbf{k}')_{nm;n'm'}
    \right],
    \label{sm_ph_spin_contracted_bse}
\end{align}
where $\alpha=0,1$ denotes spin-0 and spin-1, and $c^{p-h}_{0}=1$, $c^{p-h}_{1}=-1/3$.  Here
\begin{align}
    D^{p-h}_{\mathbf{P}}(\mathbf{k},\mathbf{k}')_{nm;n'm'}
    &=
    \sum_{\mathbf{G}}
    \lambda^{d}_{n'n}(\mathbf{k}+\mathbf{P},-\mathbf{q}-\mathbf{G})
    \lambda^{d}_{mm'}(\mathbf{k}',\mathbf{q}+\mathbf{G}),\\
    X^{p-h}_{\mathbf{P}}(\mathbf{k},\mathbf{k}')_{nm;n'm'}
    &=
    \sum_{\mathbf{G}}
    \lambda^{s}_{n'n}(\mathbf{k}+\mathbf{P},-\mathbf{q}-\mathbf{G})
    \lambda^{s}_{mm'}(\mathbf{k}',\mathbf{q}+\mathbf{G}).
    \label{sm_ph_density_spin_vertices}
\end{align}

Finally, we turn to the charge-2e particle-particle excitations. The following kernel is first written in the redundant pair basis generated by $\hat{\Delta}^\dagger_{\mathbf{P}}(\mathbf{k}) = \hat{\gamma}^\dagger_{n,s}(\mathbf{P}/2+\mathbf{k}) \hat{\gamma}^\dagger_{m,s'}(\mathbf{P}/2-\mathbf{k})$, carrying total momentum $\mathbf{P}$ and relative momentum $2\mathbf{k}$. The eigen-equation is given by:
\begin{align}
    [\hat{H}_{\text{eff}}, \hat{\Delta}^\dagger_{\mathbf{P}}(\mathbf{k})]|\Psi\rangle = \sum_{\mathbf{k}', n'm'\in\text{flat}, s''s'''} \Gamma^{p-p, \mathbf{P}}_{ns, ms'; n's'', m's'''}(\mathbf{k}, \mathbf{k}') \hat{\Delta}^\dagger_{\mathbf{P}}(\mathbf{k}')|\Psi\rangle.
\end{align}
Similar to the exciton excitation, the matrix $ \Gamma^{p-p, \mathbf{P}}_{ns, ms'; n's'', m's'''}(\mathbf{k}, \mathbf{k}')$ includes both single-particle parts and the momentum off-diagonal matrix:
\begin{equation}
    \begin{aligned} 
    \Gamma^{p-p, \mathbf{P}}_{ns, ms'; n's'', m's'''}(\mathbf{k}, \mathbf{k}') &= \delta_{\mathbf{k}\mathbf{k}'} \delta_{mm'} \delta_{s's'''} \Gamma^{p, \mathbf{P}/2+\mathbf{k}}_{ns; n's''} + \delta_{\mathbf{k}\mathbf{k}'} \delta_{nn'} \delta_{ss''} \Gamma^{p, \mathbf{P}/2-\mathbf{k}}_{ms'; m's'''} + \mathcal{V}^{p-p}_{\mathbf{P}}(\mathbf{k}, \mathbf{k}')_{ns, ms'; n's'', m's'''},\\
      \mathcal{V}^{p-p}_{\mathbf{P}}(\mathbf{k}, \mathbf{k}') &= \frac{2U}{S} \sum_{\mu=0}^3 a_\mu \sum_{\mathbf{G}} \Lambda^{(\mu)}_{n',s''; n,s}(\mathbf{P}/2+\mathbf{k}, -\mathbf{q}-\mathbf{G}) \Lambda^{(\mu)}_{m',s'''; m,s'}(\mathbf{P}/2-\mathbf{k}, \mathbf{q}+\mathbf{G}).
      \label{sm_pp_excitation}
    \end{aligned}
\end{equation}
Here the second form factor is evaluated at the initial momentum $\mathbf{P}/2-\mathbf{k}$ of the second particle, which is scattered to $\mathbf{P}/2-\mathbf{k}'$ by the momentum transfer $\mathbf{q}=\mathbf{k}-\mathbf{k}'$.  The particle-particle vertex has the opposite sign from the particle-hole vertex because both external operators are creation operators; the two equivalent commutators again give the factor of two.
The spin-0 and spin-1 particle-particle spectra are obtained by projecting the spin structure onto $(is_y)_{ss'}$ and $(is_y s^a)_{ss'}$, respectively.  Applying the same SU(2) contraction gives
\begin{align}
    \Gamma^{p-p,\alpha}_{nm;n'm'}(\mathbf{k},\mathbf{k}')
    &=
    \delta_{\mathbf{k}\mathbf{k}'}\delta_{mm'}\Gamma^{p,\mathbf{P}/2+\mathbf{k}}_{nn'}
    +
    \delta_{\mathbf{k}\mathbf{k}'}\delta_{nn'}\Gamma^{p,\mathbf{P}/2-\mathbf{k}}_{mm'}
    +\frac{2U}{S}\left[
    D^{p-p}_{\mathbf{P}}(\mathbf{k},\mathbf{k}')_{nm;n'm'}
    +c^{p-p}_{\alpha}X^{p-p}_{\mathbf{P}}(\mathbf{k},\mathbf{k}')_{nm;n'm'}
    \right],
    \label{sm_pp_spin_contracted_bse}
\end{align}
where $c^{p-p}_{0}=-1$ and $c^{p-p}_{1}=1/3$.  Here
\begin{align}
    D^{p-p}_{\mathbf{P}}(\mathbf{k},\mathbf{k}')_{nm;n'm'}
    &=
    \sum_{\mathbf{G}}
    \lambda^{d}_{n'n}\!\left(\frac{\mathbf{P}}{2}+\mathbf{k},-\mathbf{q}-\mathbf{G}\right)
    \lambda^{d}_{m'm}\!\left(\frac{\mathbf{P}}{2}-\mathbf{k},\mathbf{q}+\mathbf{G}\right),\\
    X^{p-p}_{\mathbf{P}}(\mathbf{k},\mathbf{k}')_{nm;n'm'}
    &=
    \sum_{\mathbf{G}}
    \lambda^{s}_{n'n}\!\left(\frac{\mathbf{P}}{2}+\mathbf{k},-\mathbf{q}-\mathbf{G}\right)
    \lambda^{s}_{m'm}\!\left(\frac{\mathbf{P}}{2}-\mathbf{k},\mathbf{q}+\mathbf{G}\right).
    \label{sm_pp_density_spin_vertices}
\end{align}
The physical Cooper-pair Hilbert space is obtained by imposing the fermionic exchange constraint on this spin-projected kernel.  To see this, write a charge-$2e$ eigen-operator in the redundant pair basis as
\begin{align}
    \hat{\mathcal{C}}^{2e}_{\mathbf{P}}
    =
    \sum_{\mathbf{k},n,m,s,s'}
    \Phi_{ns,ms'}(\mathbf{k})
    \hat{\gamma}^{\dagger}_{n,s}\!\left(\frac{\mathbf{P}}{2}+\mathbf{k}\right)
    \hat{\gamma}^{\dagger}_{m,s'}\!\left(\frac{\mathbf{P}}{2}-\mathbf{k}\right),
    \label{sm_pp_wavefunction_definition}
\end{align}
where $\Phi_{ns,ms'}(\mathbf{k})$ is the Cooper-pair wave function in this basis. Fermionic anti-commutation requires
\begin{align}
    \Phi_{ns,ms'}(\mathbf{k})=-\Phi_{ms',ns}(-\mathbf{k}),
    \label{sm_pp_fermion_constraint}
\end{align}
which gives
\begin{align}
    \Phi^{S=0}_{nm}(\mathbf{k})=\Phi^{S=0}_{mn}(-\mathbf{k}),\qquad
    \Phi^{S=1}_{nm}(\mathbf{k})=-\Phi^{S=1}_{mn}(-\mathbf{k})
    \label{sm_pp_spin_exchange_constraint}
\end{align}
after the spin projection.  Thus Eq.~\eqref{sm_pp_spin_contracted_bse} is diagonalized in the symmetric exchange subspace for spin singlet pairs and in the antisymmetric exchange subspace for spin triplet pairs. This projection does not discard physical Cooper-pair excitations; it removes the redundant description of the same pair under $(\mathbf{k},n,s)\leftrightarrow(-\mathbf{k},m,s')$. 
By diagonalizing the spin-resolved matrices $ \Gamma^{p-h, \mathbf{P}}_{ns, ms'; n's'', m's'''}(\mathbf{k}, \mathbf{k}'), \Gamma^{p-p, \mathbf{P}}_{ns, ms'; n's'', m's'''}(\mathbf{k}, \mathbf{k}')$ in Eq. \eqref{sm_ph_excitation} and \eqref{sm_pp_excitation}, or equivalently the SU(2)-contracted spin-projected kernels in Eq. \eqref{sm_ph_spin_contracted_bse} and Eq. \eqref{sm_pp_spin_contracted_bse} together with the pp exchange constraint in Eq. \eqref{sm_pp_spin_exchange_constraint}, we can obtain the exact two-particle excitations.

\section{E. Proof of the stiffness of QGN superconductors}
\label{sec:sm_second_order_response_stiffness}
Here we prove that, for a general class of QGN superconductors specified below, the
zero-temperature superfluid weight in the thermodynamic limit is
\begin{equation}
    D_{ij}(\nu)=\frac{1}{S}\partial_{\bar A_i}\partial_{\bar A_j}
E(\bar{\mathbf A})|_{\bar{\mathbf A}=0}=2N_{\rm flat}\nu(1-\nu)m_{\rm pair,ij}^{-1},
    \label{eq:sm_response_result_overview}
\end{equation}
 where $E(\bar{\mathbf A})$ is the ground-state energy in the presence of a
small uniform gauge field $\mathbf{A}$, and $S$ is the total area of the
system. We measure area in units of one unit-cell
area $A_{\rm uc}$, so the dimensionless area is $S=N_c$; in physical units,
$S_{\rm phys}=N_cA_{\rm uc}$. Here $N_{\rm flat}$ is
the total number of flat bands and $\nu$ is the flat-band electron filling. 
$m_{\rm pair,ij}^{-1}$ is the effective mass defined as the second-order derivatives of the exact two-particle energy $E_{pp}(\mathbf{p})$: $m_{\rm pair,ij}^{-1}=\frac{\partial^2E_{pp}[(|\mathbf b_{m,1}|/2\pi)\bar{\mathbf p}]}
{\partial \bar p_i\partial \bar p_j}|_{\bar{\mathbf p}=\bar{\mathbf Q}}$, where $\mathbf{b}_{m,1}$ the reciprocal momentum and $\bar{\mathbf p}=(2\pi/|\mathbf b_{m,1}|)\mathbf p,\bar{\mathbf Q}=(2\pi/|\mathbf b_{m,1}|)\mathbf Q$ are the dimensionless two-particle momentum and the center-of-mass Cooper pair momentum.   Hereafter, we use the unsuperscripted $D_s$ to
denote the diagonal London superfluid weight along the chosen 
twist direction $\mathbf{A}$.   


\subsection{Properties and assumptions of QGN superconductors}

We first give a brief review of basic properties of QGN models  \cite{PhysRevX.14.041004,gw85-5r92}. The interaction $\hat{V}=\sum_{\mathbf{R},\mathbf{R}^{\prime},\mu}V_{\mu}(\mathbf{R}-\mathbf{R}^{\prime})\hat{S}^{(\mu)}_{\mathbf{R}}\hat{S}^{(\mu)}_{\mathbf{R}^{\prime}}$ is positive semi-definite, and composed of Hermitian quadratic operators $\hat{S}^{(I)}_{\mathbf{R}}\sim \hat{c}^{\dagger}\hat{c}$ with $\mathbf{R}$ the label of unit-cell and $\mu$ the label of different choices of  $\hat{S}^{(\mu)}_{\mathbf{R}}$  within each unit cell. We solve the effective Hamiltonian in the flat-band space: $\hat{H}_{\text{eff}}=\hat{P}\hat{H}\hat{P}$, where $\hat{P}$ is the projection to the flat-band space.   The superconducting ground states is in the form of a generalized $\eta$-pairing state $|\eta_N\rangle\propto(\tilde\eta^\dagger)^{N/2}|0\rangle$, where
\begin{equation}
    \tilde\eta^\dagger
    =
    \sum_{\mathbf k}\sum_{m,n\in{\rm flat}}
    \hat\gamma_m^\dagger(\mathbf k+\mathbf Q/2)
    F_{mn}(\mathbf k)
    \hat\gamma_n^\dagger(-\mathbf k+\mathbf Q/2),
\end{equation}
where $m,n$ are the flat-band indices and
$F(\mathbf k)=-F^T(-\mathbf k)$. $\mathbf{Q}$ is a general center-of-mass pairing momentum and is also allowed to be zero.

We impose the following two natural conditions on QGN models that admit superconducting ground states:

(1) The superconducting ground state is unique in a fixed-electron-number sector;

(2) We require the projected form factor $F(\mathbf{k})$ is full-rank. Since it can be proved that the rank of $F(\mathbf{k})$ is independent of $\mathbf{k}$, any failure of the full-rank condition renders the model pathological, as it implies the existence of several completely empty and unpaired flat bands.

Under the above two assumptions, we can immediately arrive at the conclusion that the projected form factor $F(\mathbf{k})$ has only one nonzero singular value:
\begin{equation}
    F(\mathbf k)F^\dagger(\mathbf k)
    =F^\dagger(\mathbf k)F(\mathbf k)
    =|f|^2 I,
    \label{eq:sm_response_equal_singular_values}
\end{equation}
where $|f|$ is a momentum independent constant. This condition was likewise assumed in Ref. \cite{gw85-5r92}, where Eq. \eqref{eq:sm_response_result_overview} was conjectured to hold. Actually,if a full-rank $F(\mathbf k)$ has more than one nonzero singular values, the ground state degeneracy (in a fixed electron number sector)  diverges in the thermodynamic limit. The reason is as follows. As discussed in \cite{PhysRevX.14.041004}, if the projected form factor $F(\mathbf k)$ has more than one nonzero singular values, they must be momentum independent and $F(\mathbf{k})$ can be decomposed as: $F(\mathbf{k})=\sum^{N_{\text{sv}}}_ia_iF_i(\mathbf{k})$, where $N_{\text{sv}}$ is the number of nonzero singular values of $F(\mathbf{k})$. $a_i$ are the corresponding singular values and the singular values of $F_i(\mathbf{k})$ are only 1 or 0, and they are mutually orthogonal:
\begin{equation}
    \tr[F_i^{\dagger}(\mathbf{k})F_j(\mathbf{k})]=0,\forall i\neq j.
\end{equation}
If we denote $\tilde{\eta}^{\dagger}_i$ as the superconducting operator constructed from $F_i(\mathbf{k})$, then any state in the form $\Pi_i (\tilde{\eta}^{\dagger}_i)^{p_i}|0\rangle,\sum_i p_i=\frac{N}{2}$ is a ground state and they are mutually orthogonal, which in turn means the ground state degeneracy diverges in the thermodynamic limit.

From the property Eq. \eqref{eq:sm_response_equal_singular_values}, the normal matrix $F(\mathbf{k})$ is unitarily diagonalizable: $V^{\dagger}(\mathbf{k})F(\mathbf{k})V(\mathbf{k})=|f|\Lambda(\mathbf{k})$, where the diagonalized matrix $\Lambda_{ii}(\mathbf{k})=e^{i\theta_i(\mathbf{k})},i=1,2...N_{\text{flat}}$.  As a result,  for a system with $N_c$ unit-cells, we define the index  $l=(\mathbf{k},i)$ where $i$ labels the eigenvalues of  $F(\mathbf{k})$, and $\mathbf{k}$ takes values in half-of the Brillouin zone (here we assume $\pm\mathbf{k}$ are both allowed in the finite system Brillouin zone). As a result, there are $M=\frac{N_{\text{flat}}N_c}{2}$ allowed values of the index $l$. Then $\tilde{\eta}$ can be compactly written as:
\begin{equation}
    \tilde\eta^\dagger
    =
    \sum_{\ell=1}^{M}z_\ell P_\ell^\dagger,
    \qquad
    P_\ell^\dagger=\hat{\gamma}_{\ell 1}^\dagger
    \hat{\gamma}_{\ell 2}^\dagger ,
    \label{eq:sm_response_perfect_matching_eta}
\end{equation}
where $z_\ell=|f|e^{i\theta_i(\mathbf{k})}$, with the indices $i,\mathbf{k}$ associated with $l$ through our previous definition  $l=(\mathbf{k},i)$, and  $\hat{\gamma}_{\ell 1}^\dagger,\hat{\gamma}_{\ell 2}^\dagger$ are:
\begin{equation}
\hat{\gamma}_{\ell 1}^\dagger=\sum_m\hat{\gamma}^{\dagger}_m(\mathbf{k}+\frac{\mathbf{Q}}{2})V_{mi}(\mathbf{k}), \hat{\gamma}_{\ell 2}^\dagger=\sum_n\hat{\gamma}_n^{\dagger}(-\mathbf{k}+\frac{\mathbf{Q}}{2})V_{ni}^*(\mathbf{k}),
    \label{eq:basis}
\end{equation}
which form a complete basis of flat band fermion operators.

As a result, a normalized  superconducting ground state with $N=2p$ electrons is:
\begin{equation}
    |\eta_N\rangle
    =
    \frac{1}{\sqrt{e_p}}
    \sum_{\substack{S\subset\{1,\ldots,M\}\\ |S|=p}}
    z_S P_S^\dagger|0\rangle,
    \qquad
    e_p=|f|^{N}\binom{M}{p} ,
    \label{eq:sm_response_eta_subset}
\end{equation}
where $z_S=\prod_{\ell\in S}z_\ell$,
$P_S^\dagger=\prod_{\ell\in S}P_\ell^\dagger$.

Hereafter, we assume that the fixed-$N$ $\eta$ state is the uniquely selected
ground-state branch.  If accidental zero modes remain, the perturbation must
first be diagonalized in the zero-mode manifold and the following formulas
apply to the selected $\eta$ branch.

\subsection{Exact superfluid stiffness}
\subsubsection{General formalism}
We begin with the single-particle kinetic energy. A nonzero gauge field $\mathbf{A}$ enters into the kinetic energy through the Peierls substitution:

\begin{equation}
\begin{aligned}
    \hat H_{\rm sp}(\mathbf A)
    & =
    \sum_{\mathbf R,\mathbf R'}\sum_{\alpha\beta}
    t_{\alpha\beta,\mathbf R-\mathbf R'}
    e^{i\mathbf A\cdot
    (\mathbf R+\mathbf r_\alpha-\mathbf R'-\mathbf r_\beta)}
    \hat c_{\mathbf R\alpha}^\dagger
    \hat c_{\mathbf R'\beta}
    \\
    & =
    \sum_{\mathbf k,n}
    \epsilon_n(\mathbf k+\mathbf A)
    \hat\gamma_{\mathbf k n}^\dagger(\mathbf A)
    \hat\gamma_{\mathbf k n}(\mathbf A),
    \qquad
    \hat\gamma_{\mathbf k n}(\mathbf A)
    =
    \sum_\alpha U_{n\alpha}^\dagger(\mathbf k+\mathbf A)
    \hat c_{\mathbf k\alpha}.
\end{aligned}
    \label{eq:sm_response_microscopic_twist}
\end{equation}
Here $\alpha,\beta$ include all microscopic orbital and spin indices, and
$\mathbf r_\alpha$ denotes the position of orbital $\alpha$ within a unit
cell.  Along the principal direction
$\hat{\mathbf e}=\mathbf b_{m,1}/|\mathbf b_{m,1}|$, define the dimensionless
reciprocal-phase coordinate
\begin{equation}
    \bar A=\frac{2\pi}{|\mathbf b_{m,1}|}A,
    \qquad
    \partial_{\bar A}
    =\frac{|\mathbf b_{m,1}|}{2\pi}\partial_A .
    \label{eq:sm_response_twist_coordinate}
\end{equation}
The center-of-mass momentum of a charge-$2e$ pair changes by $2\mathbf A$, or
by $2\bar{\mathbf A}$ in this coordinate.
The effective Hamiltonian, which is the flat-band projected Hamiltonian, becomes:
\begin{equation}
\begin{aligned}
    H_{\text{eff}}(A)
    =
    &\sum_{\mu,\mathbf R,\mathbf R'}
    V_\mu(\mathbf R-\mathbf R')
    \hat P_A\hat S_{\mathbf R}^{(\mu)}(\mathbf A)\hat P_A
    \hat S_{\mathbf R'}^{(\mu)}(\mathbf A)\hat P_A
    \\
    ={}&
    \sum_{\mu,\mathbf K}V_\mu(\mathbf{K})
    \hat{\mathcal S}^{(\mu)\dagger}_A(\mathbf K)
    \hat{\mathcal S}^{(\mu)}_A(\mathbf K),
\end{aligned}
    \label{eq:sm_response_channel_form}
\end{equation}
where the quadratic $\hat S_{\mathbf R}^{(\mu)}(\mathbf A),  \hat{\mathcal S}^{(\mu)}_A(\mathbf K)$ operators are:
\begin{equation}
\begin{aligned}
    \hat S_{\mathbf R}^{(\mu)}(\mathbf A)
    & =
    \sum_{\alpha\beta}\sum_{\mathbf R_1,\mathbf R_2}
    S_{\alpha\beta}^{(\mu)}(\mathbf R_1,\mathbf R_2)
    e^{i\mathbf A\cdot
    (\mathbf R_1+\mathbf r_\alpha-\mathbf R_2-\mathbf r_\beta)}
    \hat c_{\mathbf R+\mathbf R_1,\alpha}^\dagger
    \hat c_{\mathbf R+\mathbf R_2,\beta}
    \\
    & =
    \frac{1}{N_c}
    \sum_{\alpha\beta}\sum_{\mathbf p,\mathbf q}
    e^{i\mathbf R\cdot(\mathbf p-\mathbf q)}
    S_{\alpha\beta}^{(\mu)}
    (\mathbf p+\mathbf A,\mathbf q+\mathbf A)
    \hat c_{\mathbf p\alpha}^\dagger
    \hat c_{\mathbf q\beta}.
\end{aligned}
    \label{eq:sm_response_microscopic_channel_twist}
\end{equation}
\begin{equation}
    \hat{\mathcal S}^{(\mu)}_A(\mathbf K)
    =
    \sum_{\mathbf k}
    \sum_{n,m\in\mathrm{flat}}\sum_{s,s'}
    \Lambda_{\mu,\mathbf K;ns,ms'}(\mathbf k;A)
    \hat\gamma_{n,s}^\dagger(\mathbf k+\mathbf K)
    \hat\gamma_{m,s'}(\mathbf k),
    \label{eq:sm_response_twisted_channel}
\end{equation}
where $m,n$ are the flat-band indices and $s,s^{\prime}$ are spin indices.
The projected form factor is
\begin{equation}
    \Lambda_{\mu,\mathbf K}(\mathbf k;\mathbf A)
    =
    U^\dagger(\mathbf k+\mathbf K+\mathbf A)
    S^{(\mu)}
    (\mathbf k+\mathbf K+\mathbf A,\mathbf k+\mathbf A)
    U(\mathbf k+\mathbf A),
    \label{eq:sm_response_twisted_form_factor}
\end{equation}
As proved in \hyperref[sec:sm_vanishing_projected_one_body]{Sec.~C}, the additional one-body terms generated by the remote bands cancel in the present model. This cancellation extends to a uniform charge twist in the thermodynamic limit: the shifted projectors $Q^{\pm}_{A}(\mathbf k)=Q^{\pm}(\mathbf k+\mathbf A)$ retain the chiral relation $\sigma_y Q^+_A\sigma_y=Q^-_A$, while the complete Brillouin-zone trace integrals remain invariant under a common momentum shift. After subtracting the $A$-independent remote-sector constant in this limit, Eq.~\eqref{eq:sm_response_channel_form} is the projected Hamiltonian relevant to the response.

Although we know $|\eta_N\rangle$ is the exact ground state at $\mathbf A=0$, the $ H_{\text{eff}}(\mathbf A\neq 0)$
is generally not exactly solvable.   Fortunately, we only need the second-order derivative of the ground state energy $E(\mathbf{A})$; hence, we only need the second order correction of $E(\mathbf{A})$. As a result, we expand $\hat{H}_{\text{eff}}(\mathbf{A})$ as:
\begin{equation}
    H_{\text{eff}}(\mathbf A)=H_0+A H_1+\frac{A^2}{2}H_2+O(A^3),
    \label{eq:sm_response_h_expansion}
\end{equation}
where $H_0\equiv H_{\text{eff}}(\mathbf A=0)$ and $A$ is the scalar component
along the chosen twist direction.  With $S=N_c$ in unit-cell-area units, the
finite-size London response is
\begin{equation}
    D_{\rm L}^{(N,N_c)}
    =\left(\frac{|\mathbf b_{m,1}|}{2\pi}\right)^2
    \frac{E_N''(0)}{N_c}.
    \label{eq:sm_response_london_units}
\end{equation}
For comparison between fillings, let $\nu_2=1/M$ and define the rescaled
finite-mesh estimator
\begin{equation}
    D_s^{(N,N_c)}
    =(1-\nu_2)D_{\rm L}^{(N,N_c)}
    \equiv {\cal C}\frac{E_N''(0)}{N_c},
    \qquad
    {\cal C}=(1-\nu_2)
    \left(\frac{|\mathbf b_{m,1}|}{2\pi}\right)^2.
    \label{eq:sm_response_stiffness_units}
\end{equation}
The factor $1-\nu_2$ removes the finite-size combinatorial factor associated
with the two-particle reference state.  Since $\nu_2\to0$, both responses
converge to the thermodynamic London weight $D_s$.

We further expand the ground state as:
\begin{equation}
    |\Psi_N(A)\rangle
    =
    |\eta_N\rangle+A|\chi_N\rangle+O(A^2),
    \qquad
    \langle\eta_N|\chi_N\rangle=0 .
    \label{eq:sm_response_state_expansion}
\end{equation}
Then we can get the second-order correction to the ground state energy by substituting this state into the expectation energy: 
\begin{equation}
    E_N(A)=\langle\Psi_N(A)|H_{\text{eff}}(A)|\Psi_N(A)\rangle/
\langle\Psi_N(A)|\Psi_N(A)\rangle=
    A\langle\eta_N|H_1|\eta_N\rangle
    +\frac{A^2}{2}{\cal F}[\chi_N]+O(A^3),
\end{equation}
where
\begin{equation}
    {\cal F}[\chi_N]
    =
    \langle\eta_N|H_2|\eta_N\rangle
    +2\langle\chi_N|H_0|\chi_N\rangle
    +4\, {\rm Re}\,\langle\chi_N|H_1|\eta_N\rangle .
    \label{eq:sm_response_quadratic_functional}
\end{equation}
Under the fixed-$N$ branch-selection assumption above, let
$P_0=|\eta_N\rangle\langle\eta_N|$ and $Q=1-P_0$.  The first-order correction
satisfies $|\chi_N\rangle=Q|\chi_N\rangle$.  Minimizing the quadratic
functional in this orthogonal subspace gives
 \begin{equation}
    Q\left(H_0|\chi_N\rangle+H_1|\eta_N\rangle\right)=0,
    \label{eq:variation}
\end{equation}
and hence
\begin{equation}
    |\chi_N\rangle=-H_0^+QH_1|\eta_N\rangle,
    \label{eq:sm_response_chi_solution}
\end{equation}
where $H_0^+$ is the Moore--Penrose inverse: it vanishes on $P_0$ and acts as
$1/E_n$ on every positive-energy eigenstate of $H_0$.  The second derivative
of the selected branch energy is therefore
\begin{equation}
    E_N''(0)
    =
    \langle\eta_N|H_2|\eta_N\rangle
    -
    2\langle H_1\eta_N|QH_0^+Q|H_1\eta_N\rangle .
    \label{eq:sm_response_second_order_energy}
\end{equation}
We call the first term the fixed-state curvature
$E_{\eta,N}''(0)=\langle\eta_N|H_2|\eta_N\rangle$ and denote the positive
relaxation correction by
$\Delta E_{{\rm ph},N}''=2\langle H_1\eta_N|QH_0^+Q|H_1\eta_N\rangle$.

Now in principle we have a  expression for the superfluid stiffness $D_s$,  but  it still seems to be a formidable task to extract the filling fraction $\nu$ and two-particle effective mass $m_{\text{pair}}$ from $D_s$ in a general model. In the next section, we show that the QGN  largely simplifies the problem and enables us to analytically express $D_s$ into $\nu$ and $m_{\text{pair}}$. 

\subsubsection{Simplification from QGN}
Here we first show that, in QGN models, $E_N''(0)$ arises entirely from overlaps and matrix elements involving states obtained by dressing $|\eta_N\rangle$ with particle--hole bilinears, namely $\hat O|\eta_N\rangle$, with $\hat O\sim\hat\gamma^\dagger\hat\gamma$.
We then show that the thermodynamic response obeys
$E_N''(0)=r_pE_{N=2}''(0)$, where
$r_p=p(M-p)/(M-1)$ and $p=N/2$.  On a generic finite mesh, the scalar trace
correction is retained explicitly below.

We begin with the curvature term $E_{\eta,N}''(0)=\langle\eta_N|H_2|\eta_N\rangle$. 
Differentiating $\hat{H}_{\text{eff}}(\mathbf{A})$ gives
\begin{equation}
\begin{aligned}
    H_1
    &=
    \sum_{\mu,\mathbf K}V_\mu(\mathbf{ K})
    \left(
    \dot{\hat{\mathcal S}}^{(\mu)\dagger}(\mathbf K)
    \hat{\mathcal S}^{(\mu)}(\mathbf K)
    +
    \hat{\mathcal S}^{(\mu)\dagger}(\mathbf K)
    \dot{\hat{\mathcal S}}^{(\mu)}(\mathbf K)
    \right),\\
    H_2
    &=
    \sum_{\mu,\mathbf K}V_\mu(\mathbf K)
    \left(
    \ddot{\hat{\mathcal S}}^{(\mu)\dagger}(\mathbf K)
    \hat{\mathcal S}^{(\mu)}(\mathbf K)
    +
    2\dot{\hat{\mathcal S}}^{(\mu)\dagger}(\mathbf K)
    \dot{\hat{\mathcal S}}^{(\mu)}(\mathbf K)
    +
    \hat{\mathcal S}^{(\mu)\dagger}(\mathbf K)
    \ddot{\hat{\mathcal S}}^{(\mu)}(\mathbf K)
    \right).
\end{aligned}
    \label{eq:sm_response_channel_derivatives}
\end{equation} 
with the derivatives of $\hat{S}$ operators as:
\begin{equation}
    \begin{aligned}
        \dot{\hat{\mathcal{S}}}^{(\mu)}(\mathbf{K})
    &=
    \left.\partial_A\hat{\mathcal{S}}^{(\mu)}_A(\mathbf{K})\right|_{A=0}=\sum_{\mathbf k}
    \sum_{n,m\in\mathrm{flat}}\sum_{s,s'}
     \left(\left.\partial_A\Lambda_{\mu,\mathbf K;ns,ms'}(\mathbf k;A)\right|_{A=0}\right)
    \hat\gamma_{n,s}^\dagger(\mathbf k+\mathbf K)
    \hat\gamma_{m,s'}(\mathbf k)\\
      \ddot{\hat{\mathcal{S}}}^{(\mu)}(\mathbf{K})
    &=
    \left.\partial_A^2 \hat{\mathcal{S}}^{(\mu)}_A(\mathbf{K})\right|_{A=0}=\sum_{\mathbf k}
    \sum_{n,m\in\mathrm{flat}}\sum_{s,s'}
    \left( \left.\partial_A^2\Lambda_{\mu,\mathbf K;ns,ms'}(\mathbf k;A)\right|_{A=0}\right)
    \hat\gamma_{n,s}^\dagger(\mathbf k+\mathbf K)
    \hat\gamma_{m,s'}(\mathbf k).
    \end{aligned}
\end{equation}

At zero twist, Hermiticity of the real-space channels gives
$\hat{\mathcal S}^{(\mu)\dagger}(\mathbf K)
=\hat{\mathcal S}^{(\mu)}(-\mathbf K)$.  Since the channel set is closed under
$\mathbf K\mapsto-\mathbf K$, the QGN zero-mode condition implies
\begin{equation}
    \hat{\mathcal S}^{(\mu)}(\mathbf K)|\eta_N\rangle
    =\hat{\mathcal S}^{(\mu)\dagger}(\mathbf K)|\eta_N\rangle=0.
    \label{eq:sm_response_channel_annihilation}
\end{equation}
The curvature term $E_{\eta,N}''(0)$ can therefore be expressed as
\begin{equation}
    E_{\eta,N}''(0)
    =
    \langle\eta_N|H_2|\eta_N\rangle
    =
    2\sum_{\mu,\mathbf{K}}V_\mu(\mathbf{K})
    \left\langle
    \dot{\hat{\mathcal{S}}}^{(\mu)}(\mathbf{K})\eta_N
    \middle|
    \dot{\hat{\mathcal{S}}}^{(\mu)}(\mathbf{K})\eta_N
    \right\rangle ,
    \label{eq:sm_response_fixed_eta_channel_curvature}
\end{equation}
which is exactly the sum of overlaps of states in $ {\cal V}_{\rm ph}$.

We next consider the relaxation term
$\Delta E_{{\rm ph},N}''=2\langle H_1\eta_N|QH_0^+Q|H_1\eta_N\rangle$.
Using Eq.~\eqref{eq:sm_response_channel_annihilation}, the source becomes
\begin{equation}
    H_1|\eta_N\rangle
    =
    \sum_{\mu,\mathbf{K}} V_\mu(\mathbf K)
    \hat{\mathcal{S}}^{(\mu)\dagger}(\mathbf{K})
    \dot{\hat{\mathcal{S}}}^{(\mu)}(\mathbf{K})|\eta_N\rangle
    =
    \sum_{\mu,\mathbf{K}} V_\mu (\mathbf{K})
    [\hat{\mathcal{S}}^{(\mu)\dagger}(\mathbf{K}),
    \dot{\hat{\mathcal{S}}}^{(\mu)}(\mathbf{K})]|\eta_N\rangle .
    \label{eq:sm_response_h1_eta_ph}
\end{equation}
The commutator of two particle--hole bilinear operators is again a
particle--hole bilinear operator:
\begin{equation}
    [\hat{\gamma}_a^\dagger \hat{\gamma}_b,
    \hat{\gamma}_c^\dagger \hat{\gamma}_d]
    =
    \delta_{bc}\hat{\gamma}_a^\dagger \hat{\gamma}_d
    -
    \delta_{ad}\hat{\gamma}_c^\dagger \hat{\gamma}_b ,
    \label{eq:commutator}
\end{equation}
Here $a,b=1,\ldots,2M$ label the orthonormal flat-band basis in
Eq.~\eqref{eq:basis}.  Thus $H_1|\eta_N\rangle$ is a superposition of
one-body-dressed states.  We define the relevant response space as
\begin{equation}
    {\cal V}_{\rm ph}
    =
    {\rm span}\left\{
    \hat O_X|\eta_N\rangle:
    \hat O_X=\sum_{a,b}\hat{\gamma}_a^{\dagger}X_{ab}\hat{\gamma}_b,
    \quad\operatorname{Tr}X=0
    \right\}.
    \label{eq:sm_response_ph_subspace}
\end{equation}
The equal-weight one-body density matrix gives
$\langle\eta_N|\hat O_X|\eta_N\rangle=(p/M)\operatorname{Tr}X=0$, so
$Q\hat O_X|\eta_N\rangle=\hat O_X|\eta_N\rangle$ and
$QH_1|\eta_N\rangle=H_1|\eta_N\rangle\in{\cal V}_{\rm ph}$.
Moreover, this restriction is exact because ${\cal V}_{\rm ph}$ is invariant
under $H_0$:
\begin{equation}
\begin{aligned}
    H_0\hat O_X|\eta_N\rangle
    &=\sum_{\mu,\mathbf K}V_\mu(\mathbf K)
    \hat{\mathcal S}^{(\mu)\dagger}(\mathbf K)
    [\hat{\mathcal S}^{(\mu)}(\mathbf K),\hat O_X]|\eta_N\rangle\\
    &=\sum_{\mu,\mathbf K}V_\mu(\mathbf K)
    [\hat{\mathcal S}^{(\mu)\dagger}(\mathbf K),
    [\hat{\mathcal S}^{(\mu)}(\mathbf K),\hat O_X]]|\eta_N\rangle
    \in{\cal V}_{\rm ph}.
\end{aligned}
    \label{eq:sm_response_ph_invariance}
\end{equation}
The double commutator is again a traceless one-body operator; hence the
spectral inverse $H_0^+$ preserves the same response space.

We first establish the pure $r_p$ scaling of the relaxation term.  The result
follows from a master identity for
$\langle\eta_N|\hat X^\dagger\hat Y|\eta_N\rangle$, where
$\hat X^\dagger=\sum_{ab}\hat\gamma_a^\dagger X^\dagger_{ab}\hat\gamma_b$
and $\hat Y=\sum_{ab}\hat\gamma_a^\dagger Y_{ab}\hat\gamma_b$ are one-body
bilinears:
\begin{equation}
\langle\eta_N|\hat{X}^{\dagger}\hat{Y}|\eta_N\rangle=f_p\left[\tr(X^{\dagger}Y)+\tr(X^{\dagger}JY^TJ^{\dagger})\right]+d_p\tr(X^{\dagger})\tr(Y),
\label{eq:master}
\end{equation}
where $J$ is an anti-symmetric matrix: $J_{l_1,l_2}=-J_{l_2,l_1}=e^{i\theta_i(\mathbf{k})}$ with $e^{i\theta_i(\mathbf{k})},l_1,l_2$ defined in  Eq. \eqref{eq:sm_response_perfect_matching_eta}, and all the other matrix elements of $J$ are zero. The coefficients $f_p=\frac{\binom{M-2}{p-1}}{\binom{M}{p}}
    =\frac{p(M-p)}{M(M-1)}$, which is directly related to the desired particle-number scaling factor  $r_p$ through $r_p=\frac{f_p}{f_1}$, and $d_p=\frac{p(p-1)}{M(M-1)}$. This master identity directly follows from the following three relations:
    \begin{equation}
    \begin{aligned}
         (\gamma^{\dagger}X^{\dagger}\hat{\gamma })(\gamma^{\dagger}Y\hat{\gamma })&=\hat{\gamma}^{\dagger}X^{\dagger}Y\hat{\gamma}-X_{ab}^{*}Y_{cd}\hat{\gamma}^{\dagger}_b\hat{\gamma}^{\dagger}_c
\hat{\gamma}_a \hat{\gamma}_d,\\
\langle\eta_N|\hat{\gamma}^{\dagger}_a\hat{\gamma}_b|\eta_N\rangle&=\delta_{a,b}\frac{p}{M},\\
\langle\eta_N|\hat{\gamma}^{\dagger}_a\hat{\gamma}^{\dagger}_c
\hat{\gamma}_d \hat{\gamma}_b|\eta_N\rangle&=d_p(\delta_{ab}\delta_{cd}-\delta_{ad}\delta_{cb})+f_pJ^*_{ac}J_{bd}.
    \end{aligned}
\end{equation}

To evaluate the relaxation term, choose a complete basis $\{T_\alpha\}$ of
traceless $2M\times2M$ matrices, for example
\begin{equation}
    \{T_\alpha\}
    =\{E_{ab}:a\ne b\}
    \cup\{E_{aa}-E_{2M,2M}:a=1,\ldots,2M-1\}.
    \label{eq:sm_response_complete_traceless_basis}
\end{equation}
Define $\hat O_\alpha=\hat\gamma^\dagger T_\alpha\hat\gamma$ and
$|\phi_\alpha^N\rangle=\hat O_\alpha|\eta_N\rangle$.  These states span
${\cal V}_{\rm ph}$ but can be overcomplete if the map
$T_\alpha\mapsto\hat O_\alpha|\eta_N\rangle$ has a kernel.  We define
\begin{equation}
    G^N_{\alpha\beta}=\langle\phi^N_\alpha|\phi^N_\beta\rangle,
    \qquad
    K^N_{\alpha\beta}=\langle\phi^N_\alpha|H_0|\phi^N_\beta\rangle,
    \qquad
    s^N_\alpha=\langle\phi^N_\alpha|H_1|\eta_N\rangle .
    \label{eq:sm_response_gks}
\end{equation}
Using $\hat{\mathcal S}^{(\mu)}(\mathbf K)|\eta_N\rangle=0$, these matrices
can be written in terms of traceless commutators:
\begin{equation}
    K^N_{\alpha\beta}
    =
    \sum_{\mu,\mathbf{K}} V_\mu(\mathbf{K})
    \left\langle
    [\hat{\mathcal{S}}^{(\mu)}(\mathbf{K}),\hat O_\alpha]\eta_N
    \middle|
    [\hat{\mathcal{S}}^{(\mu)}(\mathbf{K}),\hat O_\beta]\eta_N
    \right\rangle ,
    \label{eq:sm_response_k_commutator}
\end{equation}
and
\begin{equation}
    s^N_\alpha
    =
    \sum_{\mu,\mathbf{K}} V_\mu(\mathbf{K})
    \left\langle
    [\hat{\mathcal{S}}^{(\mu)}(\mathbf{K}),\hat O_\alpha]\eta_N
    \middle|
    \dot{\hat{\mathcal{S}}}^{(\mu)}(\mathbf{K})\eta_N
    \right\rangle ,
    \label{eq:sm_response_s_commutator}
\end{equation}
The commutator matrices are traceless.  Applying the master identity
Eq.~\eqref{eq:master}, including to arbitrary linear combinations of the
generators, gives
\begin{equation}
      G^N_{\alpha\beta}=r_p G^{N=2}_{\alpha\beta},\qquad
      K^N_{\alpha\beta}=r_p K^{N=2}_{\alpha\beta},\qquad
      s^N_\alpha=r_p s^{N=2}_\alpha
      \label{eq:scaling}
\end{equation}

Let the columns of $R^N$ span the non-null coefficient subspace of $G^N$ and
orthonormalize the corresponding many-body states:
\begin{equation}
    (R^N)^\dagger G^N R^N=I .
\end{equation}
For $1\le p\le M-1$, $r_p>0$, and Eq.~\eqref{eq:scaling} allows the choice
$R^N=r_p^{-1/2}R^{N=2}$.

The invariance in Eq.~\eqref{eq:sm_response_ph_invariance} ensures that the
first-order correction can be written as
$|\chi_N\rangle=\sum_\alpha(R^Ny)_\alpha|\phi_\alpha^N\rangle$.  Define
$\widetilde K^N=(R^N)^\dagger K^NR^N$ and
$\widetilde s^N=(R^N)^\dagger s^N$.  Stationarity of
Eq.~\eqref{eq:sm_response_quadratic_functional} gives
\begin{equation}
    \widetilde K^Ny_*=-\widetilde s^N,
    \qquad
    y_*=-(\widetilde K^N)^+\widetilde s^N .
\end{equation}
Substituting this solution gives
\begin{equation}
    \Delta E_{{\rm ph},N}''
    =
    2 (\widetilde s^N)^\dagger (\widetilde K^N)^+\widetilde s^N,
    \label{eq:sm_response_ph_energy}
\end{equation}
where the pseudoinverse is taken on the positive response subspace.  Combining
Eq.~\eqref{eq:scaling} with $R^N=r_p^{-1/2}R^{N=2}$ gives
\begin{equation}
     \Delta E_{{\rm ph},N}''=r_p \Delta E_{{\rm ph},N=2}''
     \label{eq:relaxation_scaling}
\end{equation}

Then we move to the curvature term $E_{\eta,N}''(0)=\langle\eta_N|H_2|\eta_N\rangle$. Using the master identity Eq.~\eqref{eq:master} to $E_{\eta,N}''(0)$, we have
\begin{equation}
    E_{\eta,N}''(0)
    =
    r_p E_{\eta,2}''(0)
    +
    2d_p\Sigma_X ,\quad \Sigma_X
    =
    \sum_{\mu,\mathbf K}V_\mu(\mathbf K)
    \left|\operatorname{Tr}\left(\partial_A\Lambda_{\mu,\mathbf K}(\mathbf k;A)|_{A=0}\right)\right|^2
    \label{eq:sm_response_fixed_eta_trace_correction}
\end{equation}
Writing $\mathbf K=\mathbf q+\mathbf G$ with $\mathbf{G}$ is the reciprocal momentum, the trace part $\operatorname{Tr}\left(\partial_A\Lambda_{\mu,\mathbf K}(\mathbf k;A)|_{A=0}\right)$ is defined as:
\begin{equation}
   \operatorname{Tr}\left(\partial_A\Lambda_{\mu,\mathbf K}(\mathbf k;A)|_{A=0}\right)
    =
    \delta_{\mathbf q,0}
    \sum_{\mathbf k,n\in\text{flat},s}
    \left.\partial_A
    \Lambda_{\mu,\mathbf G;ns,ns}(\mathbf k;A)
    \right|_{A=0}.
    \label{eq:sm_response_trace_momentum_selection}
\end{equation}
  We remark that a nonzero reciprocal momentum $\mathbf{G}$ is meaningful only in \mr models, while in finite-orbital lattice models, the reciprocal momentum $\mathbf{G}=0$. Now we show that $\frac{1}{S}\Sigma_X=0$ in the thermodynamic limit, where $S$ is the number of total unit cells, and thus $\Sigma_X$  is negligible in the curvature term $ E_{\eta,N}''(0)$, since the leading term in $ E_{\eta,N}''(0)$ should be of order $O(S)$.  This is due to the fact that the trace part $\frac{1}{S}\operatorname{Tr}\left(\partial_A\Lambda_{\mu,\mathbf K}(\mathbf k;A)|_{A=0}\right)$ is zero in the thermodynamic limit. Recalling the definition in Eq. \eqref{eq:sm_response_twisted_form_factor}, $\Lambda_{\mu,\mathbf K}(\mathbf k;A)$ is a function of $\mathbf{k}+\mathbf{A}$, so we denote $g^{\mu}_{\mathbf G}(\mathbf k+\mathbf A) =\sum_{n\in\text{flat},s}\Lambda_{\mu,\mathbf G;ns,ns}(\mathbf k;A)$. Reciprocal-lattice sewing, together with a relabeling of the plane-wave indices, gives $g^{\mu}_{\mathbf G}(\mathbf k+\mathbf H)=g^{\mu}_{\mathbf G}(\mathbf k)$ for every reciprocal vector $\mathbf H$. The Brillouin-zone integral of its momentum derivative therefore vanishes, and in the thermodynamic limit we have:
  \begin{equation}
\begin{aligned}
   \frac{1}{S}\operatorname{Tr}\left(\partial_A\Lambda_{\mu,\mathbf K}(\mathbf k;A)|_{A=0}\right)=\frac{1}{S}\delta_{\mathbf{q,0}}\partial_A\sum_{\mathbf{k}}g^{\mu}_{\mathbf G}(\mathbf k+\mathbf A)=\delta_{\mathbf{q,0}} \left.\partial_A
    \int_{\rm BZ}\frac{d^2k}{V_{\rm BZ}}
    g^{\mu}_{\mathbf G}(\mathbf k+A\hat{\mathbf e})
    \right|_{A=0}
    &=
    \int_{\rm BZ}\frac{d^2k}{V_{\rm BZ}}
    \hat{\mathbf e}\cdot\nabla_{\mathbf k}g^{\mu}_{\mathbf G}(\mathbf k)=0,
\end{aligned}
    \label{eq:sm_response_density_trace_zero}
\end{equation}
where we use the fact that $g^{\mu}_{\mathbf G}(\mathbf k)$ is periodic over the Brillouin zone and that the momentum sum becomes an integral in the thermodynamic limit.

The weight $V_\mu(\mathbf G)$ in Eq.~\eqref{eq:sm_response_channel_form} includes the $1/N_c$ Fourier normalization. Consequently,
\begin{equation}
\frac{\Sigma_X}{N_c}
=\sum_{\mu,\mathbf G}N_cV_\mu(\mathbf G)
\left|\frac{1}{N_c}\operatorname{Tr}\left(\left.\partial_A\Lambda_{\mu,\mathbf G}\right|_{A=0}\right)\right|^2
\xrightarrow{N_c\to\infty}0.
\end{equation}
This limit holds at a fixed reciprocal-lattice cutoff and extends to the sum over all reciprocal lattice vectors $\mathbf G$, provided that this sum converges uniformly in $N_c$.
In the thermodynamic limit, Eq.~\eqref{eq:sm_response_fixed_eta_trace_correction} gives the scaling relations below. At finite $N_c$, these relations hold when $\Sigma_X=0$; otherwise the trace correction must be retained.
\begin{equation}
    E_{\eta,N}''(0)
    =
    r_p E_{\eta,2}''(0).
    \label{eq:sm_response_fixed_eta_scaling}
\end{equation}

Since both the relaxation term Eq. \eqref{eq:relaxation_scaling} and the curvature term Eq. \eqref{eq:sm_response_fixed_eta_scaling} has the $r_p$ scaling, the ground state energy $E_N''(0)$ should also have the same scaling : 
\begin{equation}
    E_N''(0)
    =
    r_pE_2''(0).
\end{equation}
Using the definition of the flat-band electron filling $\nu=p/M$, the superconducting stiffness obeys:
\begin{equation}
    D_s^{(N,N_c)}
    =
    D_s^{(2,N_c)}
    \frac{\nu(1-\nu)}{\nu_2(1-\nu_2)}
    \label{eq:sm_response_n_from_n2},
\end{equation}
where $\nu_2=1/M$ represents the electron filling in the two-electron sector, and $D_s^{(2,N_c)}$ is the two-electron stiffness in a system with $N_C$ unit cells.

\subsubsection{Pair mass and thermodynamic stiffness}

Assume that the selected charge-$2e$ branch is isolated and twice
differentiable at its minimum $\mathbf Q$.
The inverse pair-mass tensor is defined from the exact two-particle dispersion (without the gauge field $\mathbf{A}=0$)
using the same reciprocal-phase coordinate as the London response.  With
$\bar{\mathbf p}=(2\pi/|\mathbf b_{m,1}|)\mathbf p$, expand near the pairing
momentum $\mathbf Q$ as
\begin{equation}
    E_{pp}\!\left(
    \mathbf Q+
    \frac{|\mathbf b_{m,1}|}{2\pi}\delta\bar{\mathbf p}
    \right)
    =
    E_{pp}(\mathbf Q)
    +
    \frac{1}{2}\delta\bar p_i
    (m_{\rm pair}^{-1})_{ij}\delta\bar p_j
    +O(|\delta\bar{\mathbf p}|^3).
    \label{eq:sm_response_pair_dispersion_mass}
\end{equation}
In the two-particle sector, it has been proved \cite{gw85-5r92} that  the second order derivative of the ground state energy $E_2(\mathbf{A})$ with respect to the gauge field $\mathbf{A}$ is proportional to $(m_{\rm pair}^{-1})_{ij}$ defined above:   $\partial_{\bar A_i}\partial_{\bar A_j}E_2(0)
=4(m_{\rm pair}^{-1})_{ij}$.
For the $C_4$-symmetric model considered here,
$m_{\rm pair}^{-1}=\frac{1}{2}\sum_i(m_{\rm pair}^{-1})_{ii}$, and in either
principal direction
\begin{equation}
    m_{\rm pair}^{-1}
    =
    \frac{1}{4}
    \left(\frac{|\mathbf{b}_{m,1}|}{2\pi}\right)^2
    E_2''(0)
    =
    \frac{D_s^{(2,N_c)}}
    {2N_{\rm flat}\nu_2(1-\nu_2)} .
    \label{eq:sm_response_mpair_from_n2_stiffness}
\end{equation}
The first equality follows from the dimensionless second-order derivative
Eq.~\eqref{eq:sm_response_twist_coordinate}.  The second equality follows from the
rescaled finite-mesh estimator in
Eq.~\eqref{eq:sm_response_stiffness_units}, because
$2N_{\rm flat}\nu_2=4/N_c$ for the two-particle state.
In the thermodynamic limit, substituting
Eq.~\eqref{eq:sm_response_mpair_from_n2_stiffness} into
Eq.~\eqref{eq:sm_response_n_from_n2} gives
\begin{equation}
    D_s(\nu)
    =
    2N_{\rm flat}\nu(1-\nu)m_{\rm pair}^{-1}.
    \label{eq:sm_response_analytic_nu_scaling}
\end{equation}


For our TBCB model, $N_{\rm flat}=4$, so this becomes
$D_s(\nu)=8\nu(1-\nu)m_{\rm pair}^{-1}$. Representative finite-mesh values are summarized in
Table~\ref{tab:sm_second_order_stiffness_finite_meshes} and plotted in the main
text.  The two-particle response fixes the pair mass, while the direct
many-body response at half filling approaches the thermodynamic scaling curve
rapidly with increasing mesh size.  The symmetry-preserving $n_k=4,6$ meshes
agree with that curve to numerical precision.

\begin{table}[t]
\centering
\begin{ruledtabular}
\begin{tabular}{c c c c c}
$n_k$ & $N_c$ & $m_{\rm pair}^{-1}$ & $D_s(N=2)$ & $D_s(\nu=1/2)$ \\
\hline
3 & 9  & 0.398870 & 0.16742680 & 0.80022112 \\
4 & 16 & 0.399819 & 0.09683112 & 0.79963772 \\
5 & 25 & 0.398758 & 0.06252520 & 0.79751792 \\
6 & 36 & 0.398845 & 0.04370060 & 0.79769000
\end{tabular}
\end{ruledtabular}
\caption{Rescaled finite-mesh second-order response of the $\eta$-paired
branch.  The
inverse pair mass is extracted from the same-mesh $N=2$ response using
Eq.~\eqref{eq:sm_response_mpair_from_n2_stiffness}.  The reported values are
the rescaled finite-mesh estimator defined in
Eq.~\eqref{eq:sm_response_stiffness_units}.}
\label{tab:sm_second_order_stiffness_finite_meshes}
\end{table}

\section{Mean-field geometric contribution and topological bound}
Having obtained the exact second-order stiffness, we next give a complementary mean-field argument for its geometric origin and topological lower bound. This section is not used in the many-body response derivation above.  Instead, it shows that a BdG description whose zero-twist ground state coincides with the exact $\eta$-paired state produces a PDW stiffness controlled by the Fubini-Study metric, in close analogy with the uniform $s$-wave pairing problem in TBG~\cite{PhysRevLett.124.167002}.

The variable $\mathbf q$ used below is the single-electron Peierls shift, equal
to $\mathbf A$ in the preceding section.  The mean-field London weight in the
same convention as the many-body response is
\begin{equation}
    [D_s^{\rm MF}]_{ij}
    =
    \left.\frac{1}{S}
    \frac{\partial^2E(\mathbf q)}{\partial q_i\partial q_j}
    \right|_{\mathbf q=0}.
    \label{eq:sm_mf_stiffness_convention}
\end{equation}
Introducing the Nambu basis
$\Psi(\mathbf{k})=\left(f_{\mathbf{Q}_l,l,\sigma,s_z}(\mathbf{k}+\frac{\mathbf{q}_1}{2}),f^{\dagger}_{\mathbf{Q}_{l^{\prime}},l^{\prime},\sigma^{\prime},s^{\prime}_z}(-\mathbf{k}+\frac{\mathbf{q}_1}{2})\right)^T$,
we adopt the following BdG mean-field approximation for a finite $\mathbf q$:
\begin{equation}
    \hat{H}_{\text{BdG}}=\frac{1}{2} \sum_{\mathbf{k}} \Psi_{\mathbf{k}}^{\dagger}\left(\begin{array}{cc}\mathcal{H}_{\text{\mr}}(\mathbf{k}+\frac{\mathbf{q}_1}{2}-\mathbf{q})-\mu & \Delta(\mathbf{k}) \\ \Delta^{\dagger}(\mathbf{k}) & -\mathcal{H}_{\text{\mr}}^{\mathrm{T}}(-\mathbf{k}+\frac{\mathbf{q}_1}{2}-\mathbf{q})+\mu\end{array}\right) \Psi_{\mathbf{k}},
\end{equation}
where $\mathbf q$ is the uniform single-fermion Peierls shift and
$\mathcal{H}_{\text{\mr}}(\mathbf{k})$ is the single-particle \mr kinetic
energy. Following the treatment in~\cite{PhysRevLett.124.167002}, we introduce the shift $\mathbf{q}$ into the kinetic energy through Peierls substitution.   The order parameter is given by:
\begin{equation}
    \Delta(\mathbf{k})=\Delta_0\tilde{U}(\mathbf{k}+\frac{\mathbf{q}_1}{2})\tilde{U}^{\dagger}(\mathbf{k}+\frac{\mathbf{q}_1}{2})\tau_xs_y\delta_{\mathbf{Q}_l,-\mathbf{Q}^{\prime}_{l^{\prime}}+\mathbf{q}_1},
\end{equation}
where  $\Delta_0$ is a constant and $\Delta(\mathbf{k})$ is the flat-band projected PDW order parameter, where the columns of $\tilde{U}(\mathbf{k})$ represent the Bloch wavefunctions of the flat bands (two flat bands per spin) at momentum $\mathbf{k}$, and $\tilde{U}(\mathbf{k})\tilde{U}^{\dagger}(\mathbf{k})$ serves as the projection operator to the flat-band space.  Transforming to the band basis where $\mathcal{H}_{\text{\mr}}(\mathbf{k})$ is diagonalized, we obtain:
\begin{equation}
    \hat{H}_{\text{BdG}}=\frac{1}{2} \sum_{\mathbf{k}} d_{\mathbf{k}}^{\dagger}\left(\begin{array}{cc}\epsilon(\mathbf{k}+\frac{\mathbf{q}_1}{2}-\mathbf{q})-\mu & U^{\dagger}(\mathbf{k}+\frac{\mathbf{q}_1}{2}-\mathbf{q})\Delta(\mathbf{k}) U^*(-\mathbf{k}+\frac{\mathbf{q}_1}{2}-\mathbf{q}) \\U^T(-\mathbf{k}+\frac{\mathbf{q}_1}{2}-\mathbf{q}) \Delta^{\dagger}(\mathbf{k}) U(\mathbf{k}+\frac{\mathbf{q}_1}{2}-\mathbf{q}) & -\epsilon(-\mathbf{k}+\frac{\mathbf{q}_1}{2}-\mathbf{q})+\mu\end{array}\right) d_{\mathbf{k}},
\end{equation}
where $d(\mathbf{k})=\left(\begin{array}{cc}U^{\dagger}(\mathbf{k}+\frac{\mathbf{q}_1}{2}-\mathbf{q}) & \\ & U^{\mathrm{T}}(-\mathbf{k}+\frac{\mathbf{q}_1}{2}-\mathbf{q})\end{array}\right) \Psi_{\mathbf{k}}$.

Next, we only focus on the flat-band space BdG Hamiltonian, since all the other bands are treated as 'frozen'. It takes the form:
\begin{equation}
\tilde{H}_{\text{BdG},\mathbf{k}}(\mathbf{q})=\left(\begin{array}{cc}-\mu & \tilde{\mathcal{D}}_{\mathbf{k}}(\mathbf{q}) \\ \tilde{\mathcal{D}}_{\mathbf{k}}^{\dagger}(\mathbf{q}) & +\mu\end{array}\right).
\end{equation}
The flat-band space BdG Hamiltonian gives the exact PDW ground state at $\mathbf q=0$.  The off-diagonal matrix is
\begin{equation}
\tilde{\mathcal{D}}_{\mathbf{k}}(\mathbf{q})=\Delta_0 \tilde{U}^{\dagger}(\mathbf{k}+\frac{\mathbf{q}_1}{2}-\mathbf{q}) \left[ \tilde{U}(\mathbf{k}+\frac{\mathbf{q}_1}{2}) \tilde{U}^{\dagger}(\mathbf{k}+\frac{\mathbf{q}_1}{2}) \right] \left(\tau_xs_y\delta_{\mathbf{Q}_l,-\mathbf{Q}_l+\mathbf{q}_1}\right) \tilde{U}^*(-\mathbf{k}+\frac{\mathbf{q}_1}{2}-\mathbf{q}).
\end{equation}
The ground state energy is:
\begin{equation}
E(\mathbf{q}) = -\frac{1}{4} \sum_{\mathbf{k}, n}\left(\sqrt{\mu^{2}+\lambda_{\mathbf{k} n}(\mathbf{q})}+\sqrt{\mu^{2}+\varphi_{\mathbf{k} n}(\mathbf{q})}\right),
\end{equation}
 where $\lambda_{\mathbf{k}n}(\mathbf{q})$ and $\varphi_{\mathbf{k} n}(\mathbf{q})$ are the eigenvalues of the positive semi-definite matrices $ \tilde{\mathcal{D}}_{\mathbf{k}}(\mathbf{q})\tilde{\mathcal{D}}_{\mathbf{k}}^{\dagger}(\mathbf{q})$ and $\tilde{\mathcal{D}}^{\dagger}_{\mathbf{k}}(\mathbf{q})\tilde{\mathcal{D}}_{\mathbf{k}}(\mathbf{q})$ .
Having obtained $E(\mathbf q)$, we obtain the mean-field London weight as
\begin{equation}
    \begin{aligned} {\left[D_s^{\rm MF}\right]_{i j}=} & \left.\frac{1}{S} \frac{\partial^{2} E(\mathbf{q})}{\partial q_{i} \partial q_{j}}\right|_{\mathbf{q}=0} \\
    = & -\frac{1}{8 S} \sum_{\mathbf{k}, n}\left(\frac{\partial_{q_{i}} \partial_{q_{j}} \lambda_{\mathbf{k} n}(\mathbf{q})}{\sqrt{\mu^{2}+\lambda_{\mathbf{k} n}(0)}}-\frac{\partial_{q_{i}} \lambda_{\mathbf{k} n}(\mathbf{q}) \partial_{q_{j}} \lambda_{\mathbf{k} n}(\mathbf{q})}{2\left[\mu^{2}+\lambda_{\mathbf{k} n}(0)\right]^{3 / 2}}\right. \left.+\frac{\partial_{q_{i}} \partial_{q_{j}} \varphi_{\mathbf{k} n}(\mathbf{q})}{\sqrt{\mu^{2}+\varphi_{\mathbf{k} n}(0)}}-\frac{\partial_{q_{i}} \varphi_{\mathbf{k} n}(\mathbf{q}) \partial_{q_{j}} \varphi_{\mathbf{k} n}(\mathbf{q})}{2\left[\mu^{2}+\varphi_{\mathbf{k} n}(0)\right]^{3 / 2}}\right)\left.\right|_{\mathbf{q}=0}\end{aligned}
\end{equation}

First,  we prove $\lambda_{\mathbf{k} n}(0)=\varphi_{\mathbf{k} n}(0)=|\Delta_0|^2$, which is a direct consequence of $\tilde{\mathcal{D}}_{\mathbf{k}}(0)\tilde{\mathcal{D}}^{\dagger}_{\mathbf{k}}(0) =\tilde{\mathcal{D}}^{\dagger}_{\mathbf{k}}(0)\tilde{\mathcal{D}}_{\mathbf{k}}(0)=|\Delta_0|^2 I$:
\begin{equation}
\begin{aligned}
\tilde{\mathcal{D}}_{\mathbf{k}}(0) &= \Delta_0 \tilde{U}^{\dagger}(\mathbf{k}+\frac{\mathbf{q}_1}{2}) \left(\tau_xs_y\delta_{\mathbf{Q}_l,-\mathbf{Q}_l+\mathbf{q}_1}\right) \tilde{U}^*(-\mathbf{k}+\frac{\mathbf{q}_1}{2}),\\
   \tilde{\mathcal{D}}_{\mathbf{k}}(0)\tilde{\mathcal{D}}^{\dagger}_{\mathbf{k}}(0)& =|\Delta_0|^2 \tilde{U}^{\dagger}(\mathbf{k}+\frac{\mathbf{q}_1}{2}) \left[\left(\tau_xs_y\delta_{\mathbf{Q}_l,-\mathbf{Q}_l+\mathbf{q}_1}\right) \tilde{U}^*(-\mathbf{k}+\frac{\mathbf{q}_1}{2}) \tilde{U}^{T}(-\mathbf{k}+\frac{\mathbf{q}_1}{2})\left(\tau_xs_y\delta_{\mathbf{Q}_l,-\mathbf{Q}_l+\mathbf{q}_1}\right)^{\dagger}\right]\tilde{U}(\mathbf{k}+\frac{\mathbf{q}_1}{2})\\
   &=|\Delta_0|^2 \tilde{U}^{\dagger}(\mathbf{k}+\frac{\mathbf{q}_1}{2}) \left[ \tilde{U}(\mathbf{k}+\frac{\mathbf{q}_1}{2})\tilde{U}^{\dagger}(\mathbf{k}+\frac{\mathbf{q}_1}{2}) \right] \tilde{U}(\mathbf{k}+\frac{\mathbf{q}_1}{2}) \\
   &= |\Delta_0|^2 I,\\
   \tilde{\mathcal{D}}^{\dagger}_{\mathbf{k}}(0)\tilde{\mathcal{D}}_{\mathbf{k}}(0)& = |\Delta_0|^2 \tilde{U}^{T}(-\mathbf{k}+\frac{\mathbf{q}_1}{2}) \left[\left(\tau_xs_y\delta_{\mathbf{Q}_l,-\mathbf{Q}_l+\mathbf{q}_1}\right)^{\dagger} \tilde{U}(\mathbf{k}+\frac{\mathbf{q}_1}{2})\tilde{U}^{\dagger}(\mathbf{k}+\frac{\mathbf{q}_1}{2})\tau_xs_y\delta_{\mathbf{Q}_l,-\mathbf{Q}_l+\mathbf{q}_1} \right] \tilde{U}^*(-\mathbf{k}+\frac{\mathbf{q}_1}{2}) \\
   &=|\Delta_0|^2 \tilde{U}^{T}(-\mathbf{k}+\frac{\mathbf{q}_1}{2}) \left[\tilde{U}^*(-\mathbf{k}+\frac{\mathbf{q}_1}{2}) \tilde{U}^{T}(-\mathbf{k}+\frac{\mathbf{q}_1}{2}) \right] \tilde{U}^*(-\mathbf{k}+\frac{\mathbf{q}_1}{2})\\
   &=|\Delta_0|^2 I.
\end{aligned}
\end{equation}
We want to remark that $\tilde{U}(\mathbf{k})$ is not a unitary matrix, since the columns of $\tilde{U}(\mathbf{k})$ only contain flat-band wave functions,  and we only have $\tilde{U}^{\dagger}(\mathbf{k})\tilde{U}(\mathbf{k})=I$ while the opposite is not true.  Besides, we have utilized the combined nonsymmorphic symmetry property $\tilde{M}_z\cdot T$: $ (u_{-\mathbf{Q}_l+l\mathbf{q}_1,-l,-s_z}^{(n)}(-\mathbf{k}+l\mathbf{q}_1))^*=e^{i\Theta_{\mathbf{k}}} u_{\mathbf{Q}_l,l,s_z}^{(n)}(\mathbf{k})$, and it gives
\begin{equation}
\left(\tau_xs_y\delta_{\mathbf{Q}_l,-\mathbf{Q}_l+\mathbf{q}_1}\right) \tilde{U}^*(-\mathbf{k}+\frac{\mathbf{q}_1}{2}-\mathbf{q}) \tilde{U}^{T}(-\mathbf{k}+\frac{\mathbf{q}_1}{2}-\mathbf{q})\left(\tau_xs_y\delta_{\mathbf{Q}_l,-\mathbf{Q}_l+\mathbf{q}_1}\right)^{\dagger}=\tilde{U}(\mathbf{k}+\frac{\mathbf{q}_1}{2}+\mathbf{q})\tilde{U}^{\dagger}(\mathbf{k}+\frac{\mathbf{q}_1}{2}+\mathbf{q}).
\label{relation}
\end{equation}

Then we prove the first-order derivatives $\partial_{q_{i}} \lambda_{\mathbf{k} n}(\mathbf{q}) =\partial_{q_{i}} \varphi_{\mathbf{k} n}(\mathbf{q})=0$. We define the following  matrix $ \mathcal{M}(\mathbf{q})$  utilizing the symmetry relation in Eq.~\eqref{relation}:
\begin{equation}
    \mathcal{M}(\mathbf{q})=\tilde{\mathcal{D}}_{\mathbf{k}}(\mathbf{q})\tilde{\mathcal{D}}^{\dagger}_{\mathbf{k}}(\mathbf{q})  = |\Delta_0|^2 A(\mathbf{q}) A^\dagger(\mathbf{q}), \quad \text{where} \quad A(\mathbf{q}) = \tilde{U}^{\dagger}(\mathbf{K}-\mathbf{q}) P(\mathbf{K}) \tilde{U}(\mathbf{K}+\mathbf{q}),
\end{equation}
with $P(\mathbf{K}) = \tilde{U}(\mathbf{K})\tilde{U}^{\dagger}(\mathbf{K})$ being the flat-band projection operator. At $\mathbf{q}=0$, $A(0) = \tilde{U}^{\dagger}(\mathbf{K}) P(\mathbf{K}) \tilde{U}(\mathbf{K}) = I$. The first-order derivative of $A(\mathbf{q})$ with respect to $\mathbf{q}$ is:
\begin{equation}
    \left. \partial_{q_i} A \right|_{\mathbf{q}=0} = -\left(\partial_{K_i} \tilde{U}^{\dagger}\right) P \tilde{U} + \tilde{U}^{\dagger} P \left(\partial_{K_i} \tilde{U}\right) = - \left(\partial_{K_i} \tilde{U}^{\dagger}\right) \tilde{U} + \tilde{U}^{\dagger} \left(\partial_{K_i} \tilde{U}\right) = 2 \tilde{U}^{\dagger} \partial_{K_i} \tilde{U}.
\end{equation}
Since $2 \tilde{U}^{\dagger} \partial_{K_i} \tilde{U}$ is purely anti-Hermitian, the first derivative of the matrix $\mathcal{M}(\mathbf{q})$ strictly vanishes: $\left. \partial_{q_i} \mathcal{M} \right|_{\mathbf{q}=0} = 0$. This in turn means $\partial_{q_{i}} \lambda_{\mathbf{k} n}(\mathbf{q}) =0$ owing to the Hellman-Feynman theorem. Similarly, we can prove $\partial_{q_{i}} \varphi_{\mathbf{k} n}(\mathbf{q})=0$.

As a result, the superfluid stiffness only contains the following terms:
\begin{equation}
   {\left[D_s^{\rm MF}\right]_{i j}=}   -\frac{1}{8 S\sqrt{\mu^2+|\Delta_0|^2}} \sum_{\mathbf{k}, n}\left(\partial_{q_{i}} \partial_{q_{j}} \lambda_{\mathbf{k} n}(\mathbf{q})+\partial_{q_{i}} \partial_{q_{j}} \varphi_{\mathbf{k} n}(\mathbf{q})\right)\left.\right|_{\mathbf{q}=0}.
\end{equation}
Now we prove that the above terms are purely quantum geometric  contributions, and we start from the second derivative of $\mathcal{M}(\mathbf{q})$:
\begin{equation}
    \operatorname{Tr} \left[ \left. \partial^2_{q_i q_j} \mathcal{M} \right|_{\mathbf{q}=0} \right] = |\Delta_0|^2 \operatorname{Tr} \left[ 2 \partial^2_{q_i q_j} A + 2 \partial_{q_i} A \partial_{q_j} A^\dagger \right]_{\mathbf{q}=0}.
\end{equation}
Using the trace identity $\operatorname{Tr}[\partial_{K_i} \tilde{U}^{\dagger} P \partial_{K_j} \tilde{U}] = - \operatorname{Tr}[\tilde{U}^{\dagger} \partial_{K_i} \tilde{U} \tilde{U}^{\dagger} \partial_{K_j} \tilde{U}]$, the second derivative term simplifies to exactly match the Fubini-Study metric:
\begin{equation}
    \operatorname{Tr} \left[ \left. \partial^2_{q_i q_j} \mathcal{M} \right|_{\mathbf{q}=0} \right] = -4 |\Delta_0|^2 \operatorname{Tr}\left[ \frac{1}{2}(\partial_{K_i} \tilde{U}^{\dagger} \partial_{K_j} \tilde{U} + \partial_{K_j} \tilde{U}^{\dagger} \partial_{K_i} \tilde{U}) + (\tilde{U}^{\dagger} \partial_{K_i} \tilde{U} \tilde{U}^{\dagger} \partial_{K_j} \tilde{U}) \right]=-4 |\Delta_0|^2  \operatorname{Tr} [g_{ij}(\mathbf{k})].
\end{equation}
Since the part $\partial_{q_{i}} \partial_{q_{j}} \varphi_{\mathbf{k} n}(\mathbf{q})$ yields an identical contribution to the stiffness, the superfluid stiffness is also proportional to the Fubini-Study metric:
\begin{equation}
    [D_s^{\rm MF}]_{ij} = -\frac{1}{4} \frac{1}{2\sqrt{\mu^2+|\Delta_0|^2}} \times 2 \times (-4 |\Delta_0|^2) \int \frac{d^{2} k}{(2 \pi)^{2}} \operatorname{Tr} [g_{ij}(\mathbf{k})] = \frac{|\Delta_0|^2}{\sqrt{\mu^2+|\Delta_0|^2}} \int \frac{d^{2} k}{(2 \pi)^{2}} \operatorname{Tr} [g_{ij}(\mathbf{k})].
\end{equation}
Finally, replacing the chemical potential $\mu$ with the flat-band electron filling fraction $\nu = \frac{1}{2}\left(1 + \frac{\mu}{\sqrt{\mu^2+|\Delta_0|^2}}\right)$ via the algebraic identity $\frac{|\Delta_0|^2}{\sqrt{\mu^2+|\Delta_0|^2}} = 2|\Delta_0|\sqrt{\nu(1-\nu)}$, we arrive at the final result:
\begin{equation}
\begin{aligned}
    [D_s^{\rm MF}]_{ij} &= 2|\Delta_0|\sqrt{\nu(1-\nu)} \int \frac{d^{2} k}{(2 \pi)^{2}} \operatorname{Tr}\left[\frac{1}{2}\left(\partial_{k_{i}} \tilde{U}_{\mathbf{k}}^{\dagger} \partial_{k_{j}} \tilde{U}_{\mathbf{k}}+\partial_{k_{j}} \tilde{U}_{\mathbf{k}}^{\dagger} \partial_{k_{i}} \tilde{U}_{\mathbf{k}}\right)+\left(\tilde{U}_{\mathbf{k}}^{\dagger} \partial_{k_{i}} \tilde{U}_{\mathbf{k}} \tilde{U}_{\mathbf{k}}^{\dagger} \partial_{k_{j}} \tilde{U}_{\mathbf{k}}\right)\right].
\end{aligned}
\end{equation}
Here the trace is over the flat-band internal indices.  The scalar stiffness quoted in the main text is $D_s^{\rm MF}=\frac{1}{2}\sum_i[D_s^{\rm MF}]_{ii}$, so the metric contribution entering the topological bound is $\sum_i\operatorname{Tr}g_{ii}(\mathbf{k})$.

\begingroup
\makeatletter
\let\@FMN@list\@empty
\makeatother
\let\supplementoriginallabel\label
\renewcommand{\label}[1]{\supplementoriginallabel{supp:#1}}
\putbib[bib]
\endgroup
\end{bibunit}
\makeatletter
\gdef\@extra@b@citeb{}
\gdef\@extra@binfo{}
\makeatother

\end{document}